\documentclass[aps,prx,twocolumn,floatfix,superscriptaddress,longbibliography]{revtex4-2}

\usepackage{graphicx}
\usepackage{amsmath,amssymb,amsthm,mathtools}
\usepackage[colorlinks=true,citecolor=blue,linkcolor=magenta]{hyperref}
\usepackage{booktabs}
\usepackage{orcidlink}
\usepackage[capitalize]{cleveref}
\usepackage{xcolor}
\usepackage{braket}

\newtheorem{theorem}{Theorem}
\newtheorem{proposition}{Proposition}
\newtheorem{lemma}{Lemma}
\newtheorem{corollary}{Corollary}

\Crefname{theorem}{Theorem}{Theorems}
\theoremstyle{remark}

\providecommand{\tr}{\mathrm{Tr}}

\newcommand{\g}{\mathfrak{g}}
\def\gsim{\g{\text - }{\rm sim}}

\newcommand{\raggedcell}[1]{{\raggedright #1\par}}

\begin{document}

\title{A Readout-Aware Theory of Approximate Heisenberg Propagation}

\author{Yubraj Bhandari\,\orcidlink{0009-0004-7279-4097}}
\affiliation{Institute for Computer Science, University of Augsburg, 86159 Augsburg, Germany}
\affiliation{Department of Physics, Duke University, Durham, North Carolina 27708, USA}

\author{Jakob~S.~Kottmann\,\orcidlink{0000-0002-4156-2048}}
\affiliation{Institute for Computer Science, University of Augsburg, 86159 Augsburg, Germany}
\affiliation{Center for Advanced Analytics and Predictive Sciences, University of Augsburg, 86159 Augsburg, Germany}

\author{Adelina Bärligea\,\orcidlink{0009-0008-5497-1941}}
\email{adelina.baerligea@uni-a.de}
\affiliation{Institute for Computer Science, University of Augsburg, 86159 Augsburg, Germany}

\begin{abstract}
Classical methods for computing quantum expectation values often propagate observables in the Heisenberg picture.
To remain tractable, these methods rely on approximations that discard operator
components during evolution. Yet the magnitude of these components alone does not determine the
resulting expectation-value error, which also depends on
their subsequent evolution and the input state.
We develop \emph{generalized Pauli propagation}, a common algebraic
framework encompassing Pauli propagation, Lie-algebraic simulation,
and Krylov subspace methods.
Using weighted commutator graphs, we derive exact identities
that relate approximation errors to the signed contributions of
omitted and discarded operator components.
Our central result establishes \emph{delayed observability},
identifying conditions under which these contributions vanish
at low orders in propagation time.
We use this structure to develop finite-order error estimates
and state-dependent selection rules.
We further prove that retaining the symmetry-preserving part
of a weak perturbation eliminates first-order expectation-value
errors under suitable symmetry conditions.
By revealing how approximation errors affect the desired
expectation value, our framework provides a basis for more
informative error analysis and the development of
readout-aware simulation strategies.
\end{abstract}

\maketitle

\section{Introduction}
\label{sec:introduction}

Classical simulation methods are central to the study of quantum
many-body dynamics, the development of quantum algorithms, and
the assessment of quantum computational advantage.
In many of these settings, the relevant task is not to
reconstruct the full quantum state but to compute selected
expectation values or correlation functions, which can remain
classically accessible even when the state itself is difficult
to represent~\cite{Bravyi.2021-ClassicalAlgorithmsQuantum}.
In the Heisenberg picture, this amounts to propagating the
measured observable instead of the state.
However, under generic dynamics, a simple observable can spread
over an exponentially large operator space.
The relevant question is therefore whether the observable can
be propagated in a tractable representation from which its
expectation value on the chosen input state can be evaluated efficiently.

Several operator-based simulation methods address this problem.
Lie-algebraic classical simulation ($\gsim$) identifies operator spaces that are invariant under commutation with the circuit generators and propagates observables exactly within them~\cite{Somma.2005-QuantumComputationComplexity,Somma.2006-EfficientSolvabilityHamiltonians,Goh.2025-LiealgebraicClassicalSimulations,Barligea.2026-EnablingLieAlgebraicClassical,Barligea.2026-LiealgebraicClassicalSimulation}. Pauli propagation expands observables in Pauli strings and evolves their coefficients directly~\cite{Rall.2019-SimulationQubitQuantum,Rudolph.2025-PauliPropagationComputational}. Majorana propagation similarly operates in a fermionic
operator basis
~\cite{Miller.2025-SimulationFermionicCircuits,
DAnna.2026-MajoranaStringSimulation,
Chakraborty.2026-Scalablequantumcircuitgeneration},
with rigorous guarantees in
weakly interacting regimes
~\cite{Facelli.2026-FastConvergenceMajorana,
Zhao.2026-EfficientClassicalSimulation}.

Although untruncated operator propagation is exact, the
number of active coefficients can grow exponentially.
This has motivated stochastic estimators
~\cite{Rall.2019-SimulationQubitQuantum},
truncated-path algorithms under noise or circuit-ensemble
assumptions
~\cite{Aharonov.2023-PolynomialTimeClassicalAlgorithm,
Shao.2025-CharacterizingPauliPropagation,
Schuster.2024-PolynomialtimeClassicalAlgorithm,
Angrisani.2025-ClassicallyEstimatingObservables,
Angrisani.2026-SimulatingQuantumCircuits},
and sparse propagation schemes that truncate by coefficient
magnitude, Pauli weight, or spatial extent
~\cite{Begusic.2025-RealTimeOperatorEvolution,
Shao.2025-CharacterizingPauliPropagation,
Rudolph.2025-PauliPropagationComputational,
HoldenDye.2026-DiameterTruncatedOperator}.
Fixed-generator Krylov methods approximate Heisenberg
propagation in subspaces generated by nested commutators
with the observable
~\cite{Saad.1992-AnalysisKrylovSubspace,
Hochbruck.1997-KrylovSubspaceApproximations,
Parker.2019-UniversalOperatorGrowth}.
Related simulation methods exploit compressibility in
tensor-network representations, which restrict the bond
dimension of a represented state or operator
~\cite{Perez-Garcia.2007-MPSrepresentations,
Markov.2008-SimulatingQCbyContractingTNs,
Hartmann.2009-DensityMatrixRenormalization,
schollwock2011density}.

These approximation methods can substantially reduce the
cost of operator propagation, but their accuracy must
ultimately be assessed through the expectation values
being computed. In particular, the magnitude of discarded
operator components alone does not determine the resulting
mean-value error. Their contribution depends on the
subsequent dynamics and on the readout defined by the
input state.

Despite their different formulations, these operator-space
methods admit a common algebraic description. Heisenberg evolution defines a linear propagation problem
on operator space, whose matrix representation is determined
by the commutator action of the Hamiltonian. Exact $\gsim$ exploits small invariant operator spaces,
Pauli propagation uses the sparse Pauli string basis, and
fixed-generator Krylov methods approximate the resulting
linear propagation in observable-generated subspaces.

This common description does not make different operator
representations computationally equivalent.
Their computational cost depends strongly on the structure
of the chosen coordinates.
A single structured
coordinate may combine exponentially many individual Pauli strings while
admitting a simpler commutator action~\cite{Barligea.2026-EnablingLieAlgebraicClassical}.
For example, symmetry-merging Pauli propagation combines symmetry-related contributions between propagation steps~\cite{Teng.2025-LeveragingSymmetryMerging}, whereas exact $\gsim$ can propagate
directly in symmetry-adapted coordinates~\cite{Barligea.2026-EnablingLieAlgebraicClassical}.

To analyze how approximation errors affect the requested
expectation value, we formulate Heisenberg evolution as
weighted source-to-readout transfer on a commutator graph,
building on existing graph descriptions of operator dynamics
~\cite{Cao.2021-StatisticalMechanismOperator,
White.2023-EffectiveDissipationRate,
West.2025-GraphTheoreticApproachChaos}.
We refer to this basis-covariant formulation as
\emph{generalized Pauli propagation}
(see \cref{sec:HeisenbergPropagation}).

Within this framework, we derive exact identities for
the mean-value error of continuous and discrete approximate
evolution (see \cref{sec:Approximation}).
These identities distinguish components omitted from the
initial observable from defects generated during propagation
and express their effects as signed contributions to the
readout. They explicitly resolve the state-dependent contributions and cancellations between discarded contributions that
are absent from norm-based error measures.

Our central structural result is a delayed-observability
theorem that identifies conditions under which the
contributions of discarded components vanish at low
orders in propagation time (see \cref{sec:Defects}).
The underlying construction combines a grading adapted
to the input-state readout with restrictions on how
commutators change that grading.
Starting from the number of $X/Y$ factors in Pauli strings
for the all-zero input state
~\cite{Begusic.2025-RealTimeOperatorEvolution},
we extend this approach to weight states and generalized
coherent states
~\cite{Perelomov.1972-CoherentStatesArbitrary,
Perelomov.1986-GeneralizedCoherentStates,
Gilmore.1972-GeometrySymmetrizedStates,
Zhang.1990-CoherentStatesTheory}.
The resulting selection rules motivate finite-order
estimates of signed readout errors.
We further establish first-order readout protection
under weak symmetry breaking when the perturbation
acting within an invariant sector is included in the approximate generator
(see \cref{subsec:WeakSymmetry}).

Our numerical studies assess the accuracy and computational cost
of signed defect-transfer estimates and demonstrate approximate
propagation directly in structured symmetry-adapted
operator coordinates. We compare the computational work with propagation
of individual operator strings and test the predicted readout protection under weak symmetry breaking (see \cref{sec:Numerics}).

Together, these results connect approximation mechanisms and
the algebraic structure of Heisenberg propagation to their
consequences for a chosen quantum expectation value.
They provide a readout-aware basis for assessing approximation
errors across different operator representations.

\section{Heisenberg propagation on commutator graphs}
\label{sec:HeisenbergPropagation}

We consider the computation of quantum mean values of the form
\begin{equation}
    \langle O(t)\rangle
    =
    \tr\!\left[
        O\,U(t)\rho_{\mathrm{in}}U^\dagger(t)
    \right]
    =
    \tr\!\left[
        U^\dagger(t) O U(t)\rho_{\mathrm{in}}
    \right],
    \label{eq:expectationvalue}
\end{equation}
where $O$ is an observable and $\rho_{\mathrm{in}}$ an input state. For $n$ qubits, the underlying Hilbert space is $\mathcal H=(\mathbb C^2)^{\otimes n}$, on which $O, \rho_{\mathrm{in}}$ and the time-evolution unitary 
$U(t)$ act. The same construction applies to any finite-dimensional invariant sector of a more general system.

\subsection{Adjoint-space propagation}
\label{subsec:AdjointSpace}

The Heisenberg form of \cref{eq:expectationvalue} identifies the dynamical object underlying the mean-value problem as the linear map that sends an operator $X$ to its conjugate $U^\dagger X U$. We formulate this map on an operator space. Let $\operatorname{Herm}(\mathcal H)$ denote the real vector space of Hermitian operators on $\mathcal H$. We consider a finite-dimensional subspace
$\mathcal V\subseteq\operatorname{Herm}(\mathcal H)$
that contains $O$ and is invariant under
the adjoint actions of the Hamiltonians considered.
For a unitary $U$ preserving $\mathcal V$, define the finite adjoint action
\begin{equation}
\operatorname{Ad}_{U}:\mathcal V\longrightarrow\mathcal V,\quad\operatorname{Ad}_{U}(X):=U^\dagger XU .
\end{equation}

For $U(t)=e^{-itH}$, conjugation is generated by the infinitesimal adjoint action on Hermitian operators,
\begin{equation}
\operatorname{ad}_H:\mathcal V\longrightarrow\mathcal V,
\quad\operatorname{ad}_H(X):=i[H,X].
\end{equation}
Indeed, this map is the tangent action of unitary conjugation, as the Heisenberg-evolved observable $O(t):=\operatorname{Ad}_{U(t)}(O)$ satisfies
\begin{equation}
    \frac{\mathrm d}{\mathrm dt}O(t)
    =
    i[H,O(t)]
    =
    \operatorname{ad}_H\!\left(O(t)\right).
    \label{eq:adjderivation}
\end{equation}
Consequently, the finite and infinitesimal actions are related by
\begin{equation}
    O(t)
    =
U^\dagger(t)OU(t)
    =
    e^{t\operatorname{ad}_H}(O),
    \label{eq:adjconjugation}
\end{equation}
or equivalently $\operatorname{Ad}_{U(t)}=e^{t\operatorname{ad}_H}$ on $\mathcal V$.

To obtain a computational representation, we now choose coordinates on $\mathcal V$.
Throughout, we equip $\mathcal V$
with the normalized Hilbert--Schmidt inner product ${\langle X,Y\rangle_\kappa
:=
\kappa\tr(XY)}$,
where $\kappa>0$ fixes the normalization; i.e., for the $n$-qubit Pauli basis,
we take $\kappa=2^{-n}$. 
Let ${\mathcal B=\{B_\alpha\}_{\alpha=1}^{D}}$ be an orthonormal Hermitian basis of $\mathcal V$, with
$D:=\dim\mathcal V$. We denote by $\Phi_{\mathcal B}^{\operatorname{ad}}(H)$ the row-action matrix representing the infinitesimal map $\operatorname{ad}_H$ in this basis. Thus, with $A:=\Phi_{\mathcal B}^{\operatorname{ad}}(H)$,
\begin{equation}
    \operatorname{ad}_H(B_\alpha)
    =
    \sum_{\beta=1}^{D}
    A_{\alpha\beta}B_\beta,
    \quad
    A_{\alpha\beta}
    =
    \left\langle
        B_\beta,\operatorname{ad}_H(B_\alpha)
    \right\rangle_\kappa.
    \label{eq:adjointmatrix}
\end{equation}
Note that if the Hamiltonian is expanded as
$H=\sum_{\mu=1}^{M}h_\mu F_\mu$
in Hermitian operators $F_\mu$, the matrix elements become
\begin{equation}
A_{\alpha\beta}
=
\sum_{\mu=1}^{M}
h_\mu
\left\langle
B_\beta,i[F_\mu,B_\alpha]
\right\rangle_\kappa.
\end{equation}
Thus, the propagation matrix is determined by commutators between
the Hamiltonian terms and the propagated basis operators.
The Hamiltonian need not belong to $\mathcal V$, so the
operators $F_\mu$ need not coincide with the basis $\mathcal B$.

Similarly, $\Phi_{\mathcal B}^{\operatorname{Ad}}(U)$ denotes the row-action matrix of finite conjugation,
\begin{equation}
U^\dagger B_\alpha U=\sum_{\beta=1}^{D}\bigl[\Phi_{\mathcal B}^{\operatorname{Ad}}(U)\bigr]_{\alpha\beta}B_\beta .
\end{equation}
The skew-adjointness of $\operatorname{ad}_H$ implies
$A^\top=-A$, and its finite action is
$\Phi_{\mathcal B}^{\operatorname{Ad}}(U(t))=e^{t\Phi_{\mathcal B}^{\operatorname{ad}}(H)}=e^{tA}$. We suppress the basis subscript below. 

Expanding the observable in the chosen operator basis and defining the input-state readout values, we obtain
\begin{equation}
    O=\sum_{\alpha=1}^{D}w_\alpha B_\alpha,
    \ w_\alpha\in\mathbb R, \quad
    s_\alpha
    :=
    \tr(B_\alpha\rho_{\mathrm{in}}).
    \label{eq:observableandinputcoordinates}
\end{equation}
The coefficient vector $w\in\mathbb R^D$, the readout vector
$s\in\mathbb R^D$, and the adjoint matrix
$A\in\mathbb R^{D\times D}$ of \cref{eq:adjointmatrix}
thus provide a finite-dimensional representation of
Heisenberg propagation and expectation-value evaluation. In particular, the observable coefficients evolve under the transpose adjoint action matrix as
\begin{equation}
    \langle O(t)\rangle
    =
    w^\top e^{tA}s
    =
    w(t)^\top s,
    \quad
    w(t)=e^{tA^\top}w.
    \label{eq:adjointmeanvalue}
\end{equation}
The quantum mean value problem~\eqref{eq:expectationvalue} has therefore been reduced to a bilinear exponential action.
Exact $\gsim$~\cite{Somma.2005-QuantumComputationComplexity,Somma.2006-EfficientSolvabilityHamiltonians,Goh.2025-LiealgebraicClassicalSimulations,Barligea.2026-EnablingLieAlgebraicClassical}, Pauli propagation~\cite{Rall.2019-SimulationQubitQuantum,Rudolph.2025-PauliPropagationComputational}, symmetry-adapted propagation~\cite{Teng.2025-LeveragingSymmetryMerging,Gargiulo.2026-PauliStringsQuantumUnified}, Majorana propagation~\cite{Miller.2025-SimulationFermionicCircuits,DAnna.2026-MajoranaStringSimulation,Chakraborty.2026-Scalablequantumcircuitgeneration}, and fixed-generator
Krylov methods~\cite{Saad.1992-AnalysisKrylovSubspace,Hochbruck.1997-KrylovSubspaceApproximations,Parker.2019-UniversalOperatorGrowth,Nandy.2025-QuantumDynamicsKrylov} evaluate this same object using different represented
operator spaces, coordinate systems, and exponential-action algorithms.
Note that the orthonormal basis is adopted here for notational clarity. The
corresponding Gram-matrix and dual-basis formulation for nonorthogonal
operator coordinates is given in
\cref{app:NonorthogonalCoordinates}.

The same formalism applies to a layered quantum circuit of the form
\begin{equation}
U(\boldsymbol{\theta})
    =
U_L\cdots U_1,\quad  U_\ell=e^{-i\theta_\ell H_\ell},
\label{eq:layeredcircuit}
\end{equation}
where $U_1$ acts first. With
$A_\ell:=\Phi^{\operatorname{ad}}(H_\ell)$, its mean value is
\begin{equation}
\langle O(\boldsymbol\theta)\rangle
=
w^\top
e^{\theta_LA_L}\cdots e^{\theta_1A_1}
s.
\label{eq:layeredadjointmeanvalue}
\end{equation}
This ordered product represents the circuit exactly. 

\subsection{Weighted commutator graph}
\label{ssec:CommGraph}
\begin{figure*}[tbph]
    \centering
    \includegraphics[width=\textwidth]{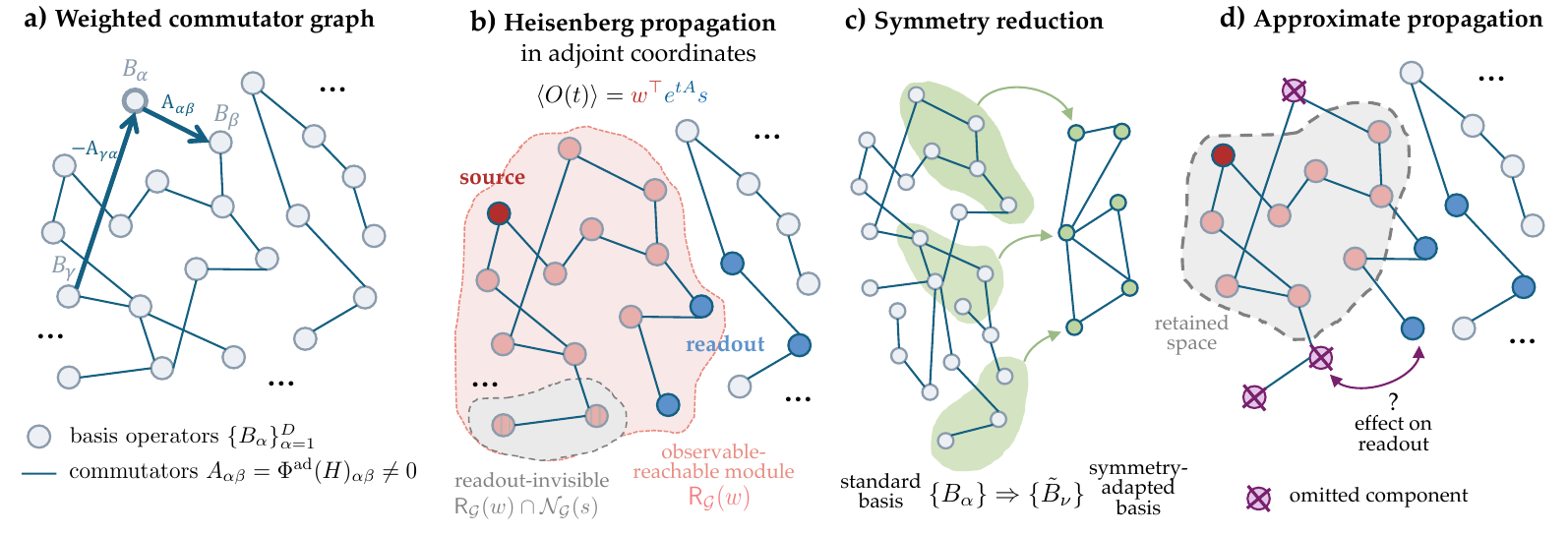}
    \caption{
        Unified graph picture of exact and approximate Heisenberg
        propagation.
        (a) In an operator basis
        $\mathcal B=\{B_\alpha\}_{\alpha=1}^{D}$, each nonzero
        row-action coefficient $A_{\alpha\beta}$ of the adjoint matrix~\eqref{eq:adjointmatrix} defines a weighted
        edge $B_\alpha\to B_\beta$ of the commutator graph, with the reverse edge carrying weight
        $-A_{\alpha\beta}$.
        (b) The mean value~\eqref{eq:adjointmeanvalue} can be interpreted as the signed transfer between
        two sets. The support of the observable coefficients $w$ defines
        the source set, while the nonzero components of
        $s$ define the state readout. Exact Heisenberg propagation requires only the
        observable-reachable module $\mathsf R_{\mathcal G}(w)$, and
        its intersection with the readout-unobservable subspace
        $\mathcal N_{\mathcal G}(s)$~\eqref{eq:unobservablesubspace} can be removed
        without affecting the computed outcome of \cref{eq:adjointmeanvalue} (cf.~\cref{prop:reachableobservable}). Note that the shaded regions are only schematic as reachable modules and
        unobservable spaces are linear subspaces and need not be spans
        of literal subsets of graph vertices.
        (c) Symmetry-related operators are combined into structured basis
elements $\{\widetilde B_\nu\}$. When their span is invariant under
the adjoint action, the resulting reduced commutator graph
describes exact propagation within that space.
        (d) Projection onto a selected operator subspace can omit initial observable components and generate boundary defects during propagation. Their contribution is determined by their subsequent transfer to the readout, as formalized in
        \cref{sec:Approximation}.
    }
    \label{fig:commutatorgraph}
\end{figure*}

The sparsity pattern of the matrix $A$ defines a directed weighted commutator graph
$\Gamma_H(\mathcal B)=(V,E,W)$. Its vertices $V=\{1,\ldots,D\}$
represent the basis operators ${B_\alpha\in\mathcal B}$, and its edges $E$ are defined via
\begin{equation}
\alpha\longrightarrow\beta
\quad\Longleftrightarrow\quad
A_{\alpha\beta}\neq0,
\quad
W(\alpha,\beta):=A_{\alpha\beta}.
\label{eq:commutatorgraph}
\end{equation}
Thus, an edge records a
nonzero component of $\operatorname{ad}_H(B_\alpha)$ along $B_\beta$.
Because $A^\top=-A$, the support graph is bidirectional, with reverse
edges carrying opposite weights. Panel~(a) of \cref{fig:commutatorgraph} illustrates this weighted
operator-basis graph. 

Related Pauli commutator graphs have been used to study operator growth~\cite{Cao.2021-StatisticalMechanismOperator}
and hydrodynamics~\cite{White.2023-EffectiveDissipationRate},
nonintegrability~\cite{Park.2025-GraphtheoreticalProofNonintegrability}, and invariant operator subspaces, scrambling, Krylov complexity, and
efficient $\gsim$-type simulation
\cite{West.2025-GraphTheoreticApproachChaos}. A
weighted commutator graph is also pruned to construct truncated shadow
Hamiltonians~\cite{Chakraborty.2026-EfficientAlgorithmApproximateShadow} for
shadow Hamiltonian simulation~\cite{Somma.2025-ShadowHamiltonianSimulation}.

Here, we equip this graph with an observable source and a state-defined
readout, turning it into a computational representation of \cref{eq:adjointmeanvalue}.
The support of the observable in $w$ (cf.~\cref{eq:observableandinputcoordinates}) defines the source set of the nodes, while the set of basis indices with nonzero input-state expectation,
\begin{equation}
S_\rho
:=
\left\{
\beta:s_\beta\neq0
\right\}
\end{equation}
defines the readout support. Expanding the matrix exponential in \cref{eq:adjointmeanvalue} gives
\begin{equation}
\langle O(t)\rangle
=
\sum_{m=0}^{\infty}
\frac{t^m}{m!}
w^\top A^m s.
\label{eq}
\end{equation}
Since $A$ is the weighted adjacency matrix of the commutator graph under
the convention of \cref{eq:commutatorgraph}, its powers encode weighted
graph walks. Concretely,
\begin{equation}
    (w^\top A^m)_\beta
    =
    \sum_{\alpha_0,\ldots,\alpha_{m-1}}
    w_{\alpha_0}
    A_{\alpha_0\alpha_1}
    A_{\alpha_1\alpha_2}
    \cdots
    A_{\alpha_{m-1}\beta},
    \label{eq:weightedwalksum}
\end{equation}
is the coherent weighted sum over
all length-$m$ walks from a source vertex $\alpha_0$ with $w_{\alpha_0}\neq0$ to the vertex $\beta$, and its weight is the source coefficient multiplied by the
edge weights along the walk. Each source-to-readout contribution is therefore weighted by its source coefficient, the
product of edge weights along the walk, the factor $t^m/m!$, and the
input-state overlap at its endpoint. 
Because these contributions add
with signs, graph connectivity determines which transfers are possible,
but not their final coefficient magnitude, as distinct walks can interfere or cancel. Panel~(b) of \cref{fig:commutatorgraph} illustrates this
source-to-readout picture.

For the layered circuit in \cref{eq:layeredadjointmeanvalue}, the vertices remain fixed, while each $A_\ell$ defines the weighted edges available during layer $\ell$.
Expanding the ordered product of exponentials produces sums over walks that traverse these edges in the prescribed layer order, through layers $L,\ldots,1$ when followed from the observable to the readout. Combining all layer edges into a single graph would ignore this ordering and therefore include paths that cannot occur in the actual circuit. 

\subsection{Exact reductions and propagation regimes}
\label{subsec:ExactPropagationRegimes}

The commutator graph describes which basis operators can couple under
the dynamics, but graph reachability need not identify the smallest
invariant operator space. For the set of circuit generators
$\mathcal G=\{H_1,\ldots,H_K\}$, this space is the
\emph{observable-reachable module}
\cite{Barligea.2026-LiealgebraicClassicalSimulation},
\begin{equation}
\begin{aligned}
\mathcal R_{\mathcal G}(O)
:=
\operatorname{span}_{\mathbb R}
\{&
\operatorname{ad}_{H_{j_m}}\cdots
\operatorname{ad}_{H_{j_1}}(O)
:\\
& m\geq0,\;
j_1,\ldots,j_m\in\{1,\ldots,K\}
\}.
\end{aligned}
\label{eq:reachopmod}
\end{equation}
It is the smallest space containing $O$ and invariant under every
generator's adjoint action. The same invariant space is termed the \emph{Dynamic Observable Subspace} in Ref.~\cite{leipold2026}. When $O$ belongs to the dynamical Lie
algebra (DLA),
\begin{equation}
\begin{aligned} 
\mathfrak{g}&=\operatorname{span}_{\mathbb{R}}\langle \mathcal G\rangle_\mathrm{Lie} & \\
&:=\operatorname{span}_{\mathbb{R}}
\Biggl \{
\begin{aligned}
&\operatorname{ad}_{H_{j_m}}\cdots
\operatorname{ad}_{H_{j_1}}\left(H_\beta\right): \\ 
 &\ m \geq 0, j_1, \ldots, j_m, \beta \in\{1, \ldots, K\}
 \end{aligned}
 \Biggr \},
\end{aligned}
\end{equation}
the latter is also invariant but can be substantially
larger than $\mathcal R_{\mathcal G}(O)$. Conversely, the span of
individually reachable graph nodes can overestimate the module because
commutator contributions may cancel. Exact $\gsim$ exploits this
invariant-space reduction
\cite{Somma.2005-QuantumComputationComplexity,Somma.2006-EfficientSolvabilityHamiltonians,Goh.2025-LiealgebraicClassicalSimulations,Barligea.2026-EnablingLieAlgebraicClassical}.
Exact polynomial-time propagation is possible when the reachable module~\eqref{eq:reachopmod} has polynomial
dimension, its basis and adjoint actions are efficiently accessible,
the required observable and input-state coordinates can be computed
efficiently, and the circuit has polynomial length (cf.~Theorem~1 of Ref.~\cite{Barligea.2026-LiealgebraicClassicalSimulation}). When symmetry-adapted operators span an invariant operator space,
the corresponding commutator graph describes exact propagation
within that space, as illustrated in
\cref{fig:commutatorgraph}(c).

The input state can permit a further exact reduction by eliminating
operator directions that remain invisible to the chosen readout.
Let
$\mathsf R_{\mathcal G}(w):=[\mathcal R_{\mathcal G}(O)]_{\mathcal B}$
denote the coordinate image of the reachable module~\eqref{eq:reachopmod}, and define the readout-unobservable subspace,
\begin{equation}
\begin{aligned}
\mathcal N_{\mathcal G}(s)
:=
\{
v:&
v^\top A_{j_1}\cdots A_{j_m}s=0
\\ & \forall \text{generator words}\ 
(j_1,\ldots,j_m)
\}.
\end{aligned}
\label{eq:unobservablesubspace}
\end{equation}
Here a generator word denotes an ordered product
$A_{j_1}\cdots A_{j_m}$, including the identity for the empty word $m=0$.

\begin{proposition}[Reachable-and-observable reduction]
\label{prop:reachableobservable}
The space $\mathcal N_{\mathcal G}(s)$ is the largest
subspace invariant under every $A_j^\top$ and contained in
$\ker (s^\top)$.
Consequently, every circuit mean value generated by
$\mathcal G$ depends on $w$ only through its equivalence class in
\begin{equation}
\mathsf R_{\mathcal G}(w)
\Big/
\left(
\mathsf R_{\mathcal G}(w)
\cap
\mathcal N_{\mathcal G}(s)
\right).
\label{eq:reachableobservablequotient}
\end{equation}
\end{proposition}

The proof is given in \cref{prf:reachableobservableproof}. 
This is the standard reachability--observability reduction of linear input--output theory~\cite{Antoulas.2005-ApproximationLargeScale}, specialized here
to adjoint propagation. Related exact operator-space reductions for
Hamiltonian and Lindblad dynamics were developed in
Ref.~\cite{Grigoletto.2025-ExactModelReduction}.

A distinct reduction arises for a fixed Hamiltonian through the
operator-Krylov construction. Under the row-action convention, the Krylov subspace generated by the first $m$ iterates is
\begin{equation}
\begin{gathered}
\mathcal K_m(A^\top,w)
=
\operatorname{span}
\left\{
w,A^\top w,\ldots,(A^\top)^{m-1}w
\right\}\\ =
\left[\operatorname{span}_{\mathbb R}
\left\{
O,\operatorname{ad}_H(O),\ldots,
\operatorname{ad}_H^{m-1}(O)
\right\}\right]_{\mathcal B},
\end{gathered}
\label{eq:observablekrylov}
\end{equation}
whose generating Krylov vectors $(A^\top)^k w$ collect the signed weighted
contributions of all length-$k$ walks from the observable support on the
commutator graph. Standard Krylov methods approximate the exponential action $e^{tA^\top}w$ within this subspace~\cite{Saad.1992-AnalysisKrylovSubspace,Hochbruck.1997-KrylovSubspaceApproximations,AlMohy.2011-ComputingActionMatrix}.
For the skew-symmetric adjoint matrix, the equivalent Hermitian
Liouvillian $M=-iA$ also permits Lanczos propagation
\cite{Parker.2019-UniversalOperatorGrowth,Nandy.2025-QuantumDynamicsKrylov}.
The exponential action becomes exact once the full cyclic space is
represented. Notice that at finite depth, Krylov approximation introduces an error
in evaluating the propagation within the represented operator space,
which is distinct from discarding operator coordinates.

Exact invariant-space reductions and finite-depth exponential-action
approximations therefore act on different aspects of the same
source-to-readout problem. Their computational benefit depends on the
dimension of the represented space, the cost of applying its adjoint
action, and the accessibility of the input-state readout.

\section{Approximate propagation as defect transfer}
\label{sec:Approximation}

The commutator-graph formulation of
\cref{sec:HeisenbergPropagation} allows us to distinguish different
ways in which approximate propagation can depart from the exact dynamics.
An initial cutoff may remove part of the observable before propagation,
while a dynamically applied cutoff discards components generated during
evolution. Other approximations modify the generator or the evaluation
of its exponential action. These mechanisms define distinct error
channels, whose effect on the requested mean value depends on how the
resulting discrepancies propagate to the input-state readout.

\subsection{Exact transfer of propagation defects}
\label{subsec:ExactDefectTransfer}

Consider a fixed Hamiltonian with adjoint action matrix $A$, and let
$\widetilde w(t)$ be any differentiable approximate observable-coefficient
trajectory embedded in the full coordinate
space for $0\leq t\leq T$. We define its initial source error and
full-space dynamic defect by
\begin{equation}
q_0
:=
w-\widetilde w(0),
\quad
d(t)
:=
A^\top\widetilde w(t)
-\dot{\widetilde w}(t),
\label{eq:defectdefinition}
\end{equation}
respectively. The source error $q_0$ measures the initial discrepancy, including
components of the observable removed by a cutoff applied before propagation. The dynamic defect $d(t)$ measures the instantaneous discrepancy
between the derivative prescribed by the exact Heisenberg equation
and the derivative followed by the approximation
(cf.~Ref.~\cite{Jawecki.2020-ComputableUpperError}). For example, restricting propagation to a selected operator space
suppresses components that the exact commutator would generate outside
that space. These suppressed contributions appear in $d(t)$ even
though they are absent from the approximate trajectory.
The accumulated coefficient error during propagation is the separate quantity
$
\epsilon_w(t)
:=
w(t)-\widetilde w(t)$.

\begin{proposition}[Continuous defect transfer]
\label{prop:continuousdefecttransfer}
For the source error and dynamic defect defined in
\cref{eq:defectdefinition}, the coefficient error satisfies
\begin{equation}
\epsilon_w(t)
=
e^{tA^\top}q_0
+
\int_0^t
e^{(t-\tau)A^\top}d(\tau)
\,\mathrm d\tau.
    \label{eq:defectduhamel}
\end{equation}
Consequently, the error in the requested mean value is
\begin{equation}
\begin{aligned}
\epsilon_\mathrm{mv}(t)
&:=
\langle O(t)\rangle
-
\widetilde w(t)^\top s
\\
&=
q_0^\top e^{tA}s
+
\int_0^t
d(\tau)^\top
e^{(t-\tau)A}s
\,\mathrm d\tau.
\end{aligned}
\label{eq:meanvaluedefecttransfer}
\end{equation}
\end{proposition}
The proof is given in
\cref{app:defecttransferproof}. 
The first term in \cref{eq:meanvaluedefecttransfer} propagates the
initial source error, while the integral accounts for defects generated
throughout the evolution. In the commutator-graph picture, each
$d(\tau)$ acts as an additional source at time $\tau$, whose
contribution is transferred to the readout by the subsequent dynamics.
The identity is exact for any differentiable approximate trajectory
and preserves the signs and cancellations between these contributions.

For a layered circuit, defects are defined by comparing each approximate
update with the corresponding exact layer action. This also covers
truncation rules that discard operator components at individual layers. Let
\begin{equation}
\mathcal E_\ell(X)
:=
U_\ell^\dagger XU_\ell
\end{equation}
be the exact Heisenberg map of layer $\ell$, and define
\begin{equation}
O_L=O,
\qquad
O_{\ell-1}
=
\mathcal E_\ell(O_\ell),
\quad
\ell=L,\ldots,1.
\end{equation}
Let $\widetilde O_L,\ldots,\widetilde O_0$ be any corresponding approximate sequence,
and define the operator error at layer $\ell$ by
\begin{equation}
\epsilon_{O,\ell}
:=
O_\ell-\widetilde O_\ell.
\label{eq:layeroperatorerror}
\end{equation}
Its source error at the start of Heisenberg propagation and its local
layer defects are
\begin{equation}
Q_L
:=
O_L-\widetilde O_L,
\qquad
D_{\ell-1}
:=
\mathcal E_\ell(\widetilde O_\ell)
-
\widetilde O_{\ell-1}.
\label{eq:generallayerdefect}
\end{equation}

\begin{proposition}[Discrete defect transfer]
\label{prop:layereddefecttransfer}
The final operator error satisfies
\begin{equation}
\epsilon_{O,0}
=
\mathcal E_1\cdots\mathcal E_L(Q_L)
+
\sum_{c=0}^{L-1}
\mathcal E_1\cdots\mathcal E_c(D_c),
\label{eq:layertelescoping}
\end{equation}
where the product is the identity for $c=0$. Consequently, the
mean-value error is
\begin{equation}
\begin{aligned}
\epsilon_{\mathrm{mv}}
&:=
\langle O_0\rangle_{\rho_{\mathrm{in}}}
-
\langle \widetilde O_0\rangle_{\rho_{\mathrm{in}}}
\\
&=
\operatorname{Tr}\!\left[
\rho_{\mathrm{in}}\,
\mathcal E_1\cdots\mathcal E_L(Q_L)
\right]
+
\sum_{c=0}^{L-1}
\operatorname{Tr}\!\left[
\rho_{\mathrm{in}}\,
\mathcal E_1\cdots\mathcal E_c(D_c)
\right].
\end{aligned}
\label{eq:layerdefecttransfer}
\end{equation}
\end{proposition}
The proof is given in
\cref{app:defecttransferproof}. 
Each local defect $D_c$ is propagated through the remaining Heisenberg
layers before contributing to the readout, while $Q_L$ accounts for
any initial source error. The identity holds for arbitrary approximate
layer updates and describes the error relative to the specified circuit,
independently of whether that circuit approximates a continuous-time
evolution.

\subsection{Projected propagation and invariant cuts}
\label{subsec:ProjectedPropagation}

A large class of truncation schemes restricts the propagated
operator to a selected subspace. Let $P=P^2$ be a
projector onto such a space and consider the projected coefficient dynamics
\begin{equation}
\widetilde w(0)=Pw,
\qquad
\dot{\widetilde w}(t)
=
PA^\top P\,\widetilde w(t).
\label{eq:projectedpropagation}
\end{equation}
For the orthonormal coordinates used in the main text, ordinary coordinate
cutoffs correspond to orthogonal projectors satisfying $P^\top=P$.
The following result requires only idempotence and therefore also applies to the
projectors discussed in Appendix~\ref{app:NonorthogonalCoordinates} that are orthogonal in a different metric.

\begin{corollary}[Source and boundary defects]
\label{cor:projecteddefect}
For the projected trajectory in
\cref{eq:projectedpropagation}, the source error and dynamic defect are
\begin{equation}
q_0=(I-P)w,
\quad
d(t)
=
(I-P)A^\top\widetilde w(t).
\label{eq:projecteddefect}
\end{equation}
\end{corollary}
The proof is given in Appendix~\ref{app:sourceboundarydefectsproof}.

\cref{eq:projecteddefect} makes the origin of a boundary
defect explicit. The exact dynamics would generate
$A^\top\widetilde w(t)$, but projected propagation includes only its
component inside the selected space. The discarded component
$(I-P)A^\top\widetilde w(t)$ is the dynamic boundary defect, as illustrated in \cref{fig:commutatorgraph}(d). 
For a coordinate cutoff, it collects the weighted couplings from
vertices of the commutator graph inside the cutoff to vertices outside it. For example,
under a Pauli-weight cutoff, commutators of strings within the cutoff can
generate strings above the cutoff, whose contributions are then
discarded. If $Pw=w$, the initial source error vanishes, but such
boundary defects may still arise. However, their magnitude alone does not
determine the mean-value error, which depends on their subsequent
transfer to the readout.

The layered analogue is immediate. For a projected update
\begin{equation}
\widetilde O_{\ell-1}
=
P_{\ell-1}\mathcal E_\ell(\widetilde O_\ell),
\end{equation}
the local defect of \cref{eq:generallayerdefect} is precisely the
discarded boundary component,
\begin{equation}
D_{\ell-1}
=
(I-P_{\ell-1})
\mathcal E_\ell(\widetilde O_\ell).
\label{eq:projectedlayerdefect}
\end{equation}
Thus, the continuous defect $d(t)$~\eqref{eq:defectdefinition} describes the discarded contribution
to the instantaneous derivative, while $D_{\ell-1}$~\eqref{eq:projectedlayerdefect} is the component
discarded by a complete layer update. Both arise when the exact
propagation leaves this subspace.

Invariant-sector truncation is the limiting case in which the retained ($\operatorname{ran}P$)
and discarded ($\ker P$) spaces are dynamically decoupled, i.e., evolve independently. If $[A^\top,P]=0$, the dynamic defect $d(t)$ vanishes, and the only error comes from the initially omitted part $(I-P)w$ of the observable. 
Its contribution to the mean value is
\begin{equation}
\epsilon_\mathrm{mv}(t)
=
w^\top(I-P)^\top e^{tA}s.
\label{eq:invarianttail}
\end{equation}
This error vanishes for all $t$ if the omitted component is readout-unobservable in the sense of \cref{prop:reachableobservable}. A sufficient condition for this is 
$(I-P)^\top s=0$, meaning that the state readout annihilates the discarded space. This idea includes the invariant-irrep truncation of
Ref.~\cite{Monbroussou.2026-ClassicalSimulationModel}, which discards complete invariant sectors of a group representation. Appendix~\ref{app:sharedirreps} relates this invariant-sector limit
to the shared support of the observable and state readout.

A conservative bound follows from \cref{eq:meanvaluedefecttransfer}
and the orthogonality of $e^{tA}$, which holds because $A^\top=-A$
in the orthonormal coordinates used here,
\begin{equation}
|\epsilon_\mathrm{mv}(t)|
\leq
\|s\|_2
\left(
\|q_0\|_2
+
\int_0^t
\|d(\tau)\|_2
\,\mathrm d\tau
\right).
\label{eq:globaldefectbound}
\end{equation}
While general, the bound uses only the source-error and defect norms together with
$\|s\|_2$, so it does not resolve the individual signed contributions
or their cancellations in \cref{eq:meanvaluedefecttransfer}.

Finally, the same defect-transfer identities apply beyond projected propagation,
including modified generators and approximate exponential actions.
Appendix~\ref{app:TruncationAtlas} compares representative simulation
methods in this framework and distinguishes errors in the propagated
observable from errors introduced when estimating their transfer
to the readout.

\section{Delayed observability and readout-error estimates}
\label{sec:Defects}

The exact identities of \cref{sec:Approximation} reduce the effect of a
discarded component to its subsequent transfer to the input-state readout.
We now characterize when the commutator-graph structure delays this transfer
and how that delay can be used for readout-aware defect estimation.
Panel~(a) of \cref{fig:defectreadout} illustrates the corresponding
defect-to-readout picture.

\subsection{Readout grading and delayed observability}
\label{subsec:DelayedObservability}

To quantify how far an operator component is from contributing to the
chosen readout, we assign each basis operator a nonnegative integer \emph{readout grade}, chosen so that positive-grade operators have zero expectation in the input state. Grade-zero operators may have nonzero expectation. If a single commutator can lower the grade only by a bounded amount,
the grade provides a lower bound on the number of commutator steps
required before a positive-grade component can contribute to the readout.

The simplest example is
$\rho_{\mathrm{in}}=\lvert0\rangle\!\langle0\rvert^{\otimes n}$ in the
Hermitian Pauli basis $\mathcal P_n$. Since
\begin{equation}
\rho_{\mathrm{in}}
=
\frac{1}{2^n}
\prod_{j=1}^n(I+Z_j),
\end{equation}
a Pauli string has nonzero input expectation only if it contains no
$X$ or $Y$ factors. The number of $X/Y$ factors therefore
defines the natural state-dependent grade
\begin{equation}
    \operatorname{gr}_{XY}(P)
    :=
    \left|
        \left\{
            j:P_j\in\{X,Y\}
        \right\}
    \right|.
    \label{eq:xygrade}
\end{equation}
Thus, $\operatorname{gr}_{XY}(P)=0$ precisely for the readout-visible
$I/Z$ sector, while increasing grade counts how many local factors are
invisible to the input state before the dynamics acts. This quantity has
previously been used as a practical truncation criterion in sparse Pauli
evolution~\cite{Begusic.2025-RealTimeOperatorEvolution}. Here, its role is
instead to organize operator directions by their possible delay before
becoming visible to the readout. For example,
$\operatorname{gr}_{XY}(Z^{\otimes n})=0$ even though the ordinary Pauli
support weight of $Z^{\otimes n}$ is $n$.

The grading argument of \cref{eq:xygrade} extends to arbitrary operator bases. We define the readout grading described above as a function $\operatorname{gr}:\mathcal B\longrightarrow\mathbb Z_{\geq0}$. The grading is
chosen relative to the input state such that operators of positive
grade have vanishing expectation values,
\begin{equation}
    \operatorname{gr}(B_\alpha)>0 
    \quad\Longrightarrow\quad
    s_\alpha=0
    \label{eq:gradereadoutcondition}
\end{equation}
Suppose that one adjoint step can lower the grade by at most $\delta_{\mathrm{gr}}\geq0$,
\begin{equation}
    A_{\alpha\beta}\neq0
    \quad\Longrightarrow\quad
    \operatorname{gr}(B_\beta)
    \geq
    \operatorname{gr}(B_\alpha)-\delta_{\mathrm{gr}}.
    \label{eq:gradeloweringcondition}
\end{equation}
This means that each commutator-graph edge can move a component toward the
readout-visible sector by at most $\delta_{\mathrm{gr}}$ grades. For a
nonzero defect coefficient vector ${r=(r_\alpha)_{\alpha=1}^D}$, its minimum supported grade is
\begin{equation}
    \operatorname{gr}_{\min}(r)
    :=
    \min_{r_\alpha\neq0}
    \operatorname{gr}(B_\alpha).
    \label{eq:minimumdefectgrade}
\end{equation}
To quantify how a defect affects the readout, we define its
\emph{transfer function}
\begin{equation}
\Phi_r(t)
:=
r^\top e^{tA}s,
\label{eq:defecttransferfunction}
\end{equation}
which gives the signed contribution of a defect $r$ after a
further propagation time $t$.
Expanding the exponential yields
\begin{equation}
\Phi_r(t)
=
\sum_{m=0}^{\infty}
\frac{t^m}{m!}\mu_m(r),
\qquad
\mu_m(r)
:=
r^\top A^m s.
\label{eq:transfermoments}
\end{equation}
We call $\mu_m(r)$ the $m$-th \emph{transfer moment}. It is the
$m$-th derivative of $\Phi_r(t)$ at $t=0$ and equals the signed
sum of weighted commutator-graph walks of length $m$ from the
defect to the readout. Thus, the transfer moments determine
the short-time behavior of the defect's contribution to the
mean-value error.

\begin{theorem}[Delayed observability]
\label{thm:delayedobservability}
Suppose \cref{eq:gradereadoutcondition,eq:gradeloweringcondition}
hold with $\delta_{\mathrm{gr}}\geq0$.
For a nonzero defect $r$ and $\delta_{\mathrm{gr}}>0$, define
\begin{equation}
    \Delta(r)
    :=
    \left\lceil
        \frac{\operatorname{gr}_{\min}(r)}{\delta_{\mathrm{gr}}}
    \right\rceil.
    \label{eq:defectdelay}
\end{equation}
Then all transfer moments~\eqref{eq:transfermoments} of order below $\Delta(r)$ vanish,
\begin{equation}
\mu_m(r)
=
0,
\qquad
0\leq m<\Delta(r).
\label{eq:forbiddenmoments}
\end{equation}
Consequently, the defect transfer function satisfies
\begin{equation}
\Phi_r(t)
=
\sum_{m=\Delta(r)}^\infty
\frac{t^m}{m!}\mu_m(r).
\label{eq:delayedtransferexpansion}
\end{equation}
If $\delta_{\mathrm{gr}}=0$ and
$\operatorname{gr}_{\min}(r)>0$, all transfer moments vanish and $\Phi_r(t)=0$ for every $t$.
\end{theorem}

\begin{figure*}[htbp]
    \centering
    \includegraphics[width=\textwidth]{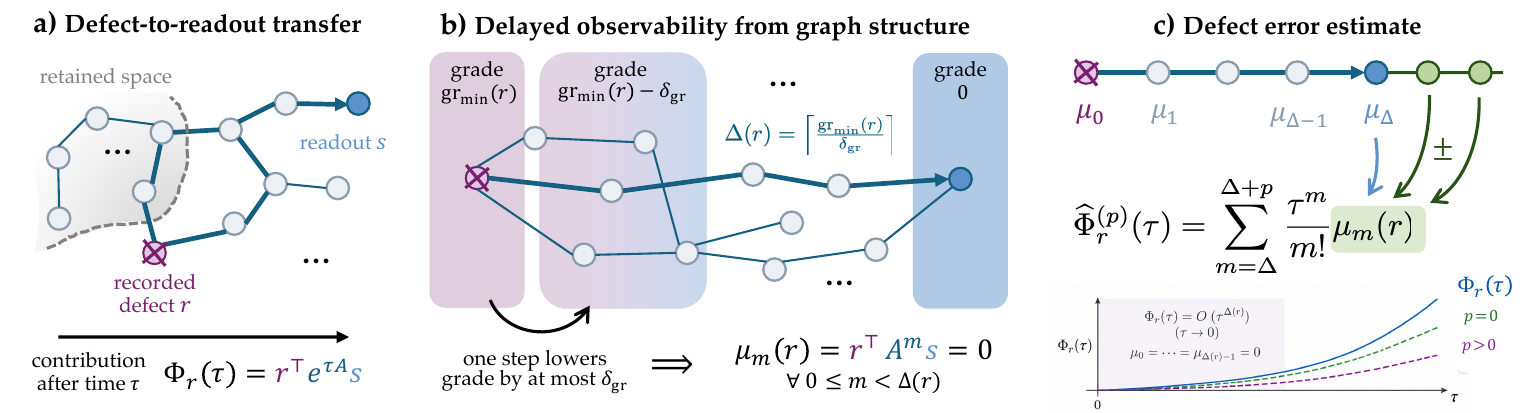}
    \caption{
Delayed observability and readout-aware defect error estimation.
(a) A recorded defect $r$ contributes to the readout through its
transfer function $\Phi_r(t)$~\eqref{eq:defecttransferfunction}, which depends on the remaining
propagation time.
(b) The readout grading excludes contributions from graph walks
shorter than $\Delta(r)$~\eqref{eq:defectdelay} steps. Consequently, the corresponding
transfer moments $\mu_m(r)$~\eqref{eq:transfermoments} vanish for $m<\Delta(r)$, as established
in \cref{thm:delayedobservability}.
(c) These vanishing moments motivate the signed error estimates
of \cref{eq:laoestimate,eq:enrichedestimate}. The leading-allowed-order
estimate uses the first allowed moment, while the $p$-enriched
estimate includes the next $p$ moments. Both approximate the exact
defect transfer through a truncated expansion in the remaining
propagation time.
}
    \label{fig:defectreadout}
\end{figure*}

The proof is given in Appendix~\ref{app:delayedobsproof}.
The integer $\Delta(r)$ is the first transfer order
not excluded by the grading conditions. In the commutator-graph
picture, it gives a lower bound on the number of steps required
for a defect to reach the readout-visible sector, as illustrated
in panel~(b) of \cref{fig:defectreadout}.
For example, if $\operatorname{gr}_{\min}(r)=3$ and
$\delta_{\mathrm{gr}}=1$, the moments
$\mu_0(r)$, $\mu_1(r)$, and $\mu_2(r)$ vanish exactly.
The earliest possible contribution is then proportional to
$t^3$, although cancellations or additional selection rules
may also eliminate this term.
Thus, $\Delta(r)$ is a lower bound on the first nonzero
transfer order, which may be larger or may not exist.

For layered dynamics, successive commutator steps may involve
different circuit generators. If every generator
$H_j\in\mathcal G=\{H_1,\ldots,H_K\}$ satisfies the same
grade-lowering bound, the delayed-observability result
extends to arbitrary generator sequences,
\begin{equation}
r^\top A_{j_1}\cdots A_{j_m}s=0
\qquad
\text{for }m<\Delta(r).
\label{eq:forbiddengeneratorwords}
\end{equation}
The relation between vanishing low-order transfer moments and
exact readout unobservability is detailed in
Appendix~\ref{app:FiniteWordUnobservability}.

\subsection{Weight-adapted gradings and generalized coherent states}
\label{subsec:StateAdaptedGradings}

An explicit readout grading can be constructed from the algebraic structure
of the input state. Consider a representation of a compact Lie
algebra on $\mathcal H$, and choose linearly independent, commuting
Hermitian generators $h_1,\ldots,h_r$ spanning a Cartan subalgebra
$\mathfrak h$, with $r=\dim\mathfrak h$. A weight state is a simultaneous eigenstate of these generators.
We use the corresponding operator weights~\cite{Hall.2015-LieGroupsLie} to identify which
operator components are invisible to its readout.
Further details and examples of these weight-based readouts are given in Appendix~\ref{app:CartanPrimer}.

Let $\mathcal V_{\mathbb C}$ be the complexification of the
Hermitian operator space $\mathcal V$, and suppose that the
commutator actions of the $h_k$ preserve $\mathcal V_{\mathbb C}$.
Since the actions $\operatorname{ad}_{h_k}$ commute, the space decomposes
into simultaneous eigenspaces
\begin{equation}
\begin{gathered}
\mathcal V_{\mathbb C}
=
\bigoplus_{\nu\in\Lambda}\mathcal V_\nu,
\\
\mathcal V_\nu
:=
\left\{
X\in\mathcal V_{\mathbb C}:
[h_k,X]=\nu_k X
\quad \forall\,k=1,\ldots,r
\right\},
\end{gathered}
\label{eq:operatorweightdecomposition}
\end{equation}
where $\nu=(\nu_1,\ldots,\nu_r)$ is an operator weight and
$\Lambda$ is the set of weights appearing in the representation.
In particular, $\mathcal V_0$ contains the operators that commute
with every $h_k\in\mathfrak h$.

\begin{proposition}[Weight-space readout selection]
\label{prop:cartanreadout}
If the input state satisfies
\begin{equation}
    [\rho_{\mathrm{in}},h_k]=0
    \quad
    \forall k,
    \label{eq:cartandiagonalstate}
\end{equation}
then every operator of nonzero weight has vanishing expectation,
\begin{equation}
    \operatorname{Tr}(\rho_{\mathrm{in}}X_\nu)
    =
    0
    \quad
    \forall \ \nu\neq0.
    \label{eq:weightselectionrule}
\end{equation}
\end{proposition}

The proof is given in Appendix~\ref{app:WeightSelectionProofs}. 
Every pure weight state satisfies \cref{eq:cartandiagonalstate},
although the proposition also applies to mixed states commuting
with the Cartan generators. Thus, the zero-weight sector is the
only sector that can contribute directly to the readout (cf.~\cref{thm:delayedobservability}).

To obtain a grading for delayed observability, we must also bound
how far one commutator can move an operator between weight sectors.
For this, decompose the Hamiltonian into components of definite weight,
\begin{equation}
H=\sum_{\mu\in\Lambda_H}H_\mu,
\qquad
[h_k,H_\mu]=\mu_k H_\mu,
\label{eq:hamiltonianweights}
\end{equation}
where $\Lambda_H$ denotes the Hamiltonian weights. The components
$H_\mu$ need not individually be Hermitian or belong to
$\mathcal V_{\mathbb C}$, but their commutator actions are assumed
to preserve this space.

Consider an integer-valued length $\ell$ on
the relevant weight lattice satisfying
\begin{equation}
\begin{gathered}
\ell(0)=0,\quad
    \ell(-\nu)=\ell(\nu)>0\ \text{for }\nu\neq0,
    \\
    \ell(\nu+\mu)\geq\ell(\nu)-\ell(\mu),
\end{gathered}
    \label{eq:weightlengthcondition}
\end{equation}
and assign each weight sector the grade
\begin{equation}
    \operatorname{gr}(X_\nu):=\ell(\nu),
    \quad
    \delta_{\mathrm{gr}}
:=
\max_{\mu\in\Lambda_H}\ell(\mu)
    \label{eq:weightgrade}
\end{equation}
for nonzero $X_\nu\in\mathcal V_\nu$.
The symmetry $\ell(-\nu)=\ell(\nu)$ allows the resulting
graded subspaces to be represented by Hermitian basis operators,
as required by \cref{thm:delayedobservability}.

\begin{lemma}[Operator-weight shift and grade jump]
\label{lem:weightshift}
Let $H_\mu$ be a definite-weight Hamiltonian component
whose commutator action preserves $\mathcal V_{\mathbb C}$.
For $X_\nu\in\mathcal V_\nu$, every nonzero commutator
satisfies
\begin{equation}
[H_\mu,X_\nu]\in\mathcal V_{\mu+\nu}.
\end{equation}
Consequently, one application of $\operatorname{ad}_H$
can lower the grade by at most $\delta_{\mathrm{gr}}$.
\end{lemma}

The proof is given in \cref{app:WeightSelectionProofs}.
Together with \cref{prop:cartanreadout}, this establishes the
conditions of \cref{thm:delayedobservability} in a basis adapted
to the operator-weight grading.

For example, for the all-zero qubit state, the commuting generators
$\{Z_1,\ldots,Z_n\}$ give the local
weight basis $\{I,Z,\sigma^+,\sigma^-\}$, where
$\sigma^\pm=(X\pm iY)/2$. Counting nonzero local weights reproduces
the $X/Y$ grading of \cref{eq:xygrade} when the corresponding equal-grade weight components are expressed in Hermitian Pauli
coordinates. Thus the preceding construction directly generalizes the
Pauli example, with details given in
Appendix~\ref{app:PauliWeightSpecialization}.

The construction also applies to generalized coherent states (GCSs). These are obtained by acting with the corresponding Lie group on
an extremal weight state
$\lvert\lambda\rangle$~\cite{Perelomov.1972-CoherentStatesArbitrary,Gilmore.1972-GeometrySymmetrizedStates,Perelomov.1986-GeneralizedCoherentStates,Zhang.1990-CoherentStatesTheory}. For $\lvert\psi\rangle=U\lvert\lambda\rangle$, the density
operator $\lvert\psi\rangle\!\langle\psi\rvert$ commutes with the
rotated Cartan algebra
$\mathfrak h_U=U\mathfrak h U^\dagger$. Such GCS inputs also occur in the original Lie-algebraic simulation
constructions
\cite{Somma.2005-QuantumComputationComplexity,
Somma.2006-EfficientSolvabilityHamiltonians}.

\subsection{Finite-order readout-transfer estimates}
\label{subsec:ReadoutTransferEstimators}

The delayed observability of \cref{thm:delayedobservability} identifies the transfer moments that vanish exactly. We use this information to approximate the
contribution of a recorded defect to the requested mean value.
For a defect $r$ with finite $\Delta(r)$, the expansion of its
transfer function $\Phi_r(\tau)$ has no terms below order $\Delta(r)$. Using only the first
allowed term gives the leading-allowed-order estimate
\begin{equation}
\widehat\Phi_r^{(0)}(\tau)
=
\frac{\tau^{\Delta(r)}}{\Delta(r)!}
\mu_{\Delta(r)}(r).
\label{eq:laoestimate}
\end{equation}
Including the next $p$ transfer moments gives the
$p$-enriched estimate
\begin{equation}
\widehat\Phi_r^{(p)}(\tau)
=
\sum_{m=\Delta(r)}^{\Delta(r)+p}
\frac{\tau^m}{m!}
\mu_m(r).
\label{eq:enrichedestimate}
\end{equation}
Here $\tau$ is the remaining propagation time after the defect
was generated. The estimates are illustrated in panel~(c) of
\cref{fig:defectreadout}.

These expressions estimate the signed contribution of a recorded
defect to a particular readout. They are therefore complementary to
norm-based truncation controls
\cite{Shao.2025-CharacterizingPauliPropagation}. If the defect contains components of different grades, these may
be treated separately to exploit their individual first allowed
orders. Such a decomposition is not required for the estimates.

The same expressions apply to the initial source error $q_0$ of \cref{eq:defectdefinition}.
For continuous propagation, replacing
$\Phi_{d(\tau)}(t-\tau)$ by its finite-order estimate in
\cref{eq:meanvaluedefecttransfer} gives an approximation to the
accumulated dynamic error.

For a layered circuit, let $d_c$ denote the coefficient vector
of the local defect $D_c$ defined in
\cref{eq:generallayerdefect}. Its exact contribution after the
remaining Heisenberg layers is
\begin{equation}
\Phi_c(d_c)
=
d_c^\top
e^{\theta_cA_c}\cdots e^{\theta_1A_1}
s,
\label{eq:ordereddefecttransfer}
\end{equation}
where the product is the identity for $c=0$.
Expanding the ordered product gives analogous finite-order
estimates, with the forbidden low-order terms absent when the generators satisfy the common grading condition of \cref{eq:forbiddengeneratorwords}. The continuous convolution, ordered layered
expansion, and conservative estimates are given in
Appendix~\ref{app:ReadoutEstimator}.

Finally, the usefulness of these estimates depends on the cost of computing the transfer moments and the size of the neglected higher-order terms. A
larger $\Delta(r)$ does not by itself guarantee a more accurate fixed-order estimate or a smaller computational cost. For longer remaining propagation times, a separate Krylov evaluation
of the transfer function may be preferable, with its own
exponential-action error
\cite{Saad.1992-AnalysisKrylovSubspace,
Jawecki.2020-ComputableUpperError}.

\subsection{Readout protection under weak symmetry breaking}
\label{subsec:WeakSymmetry}

The source-to-readout perspective also reveals a perturbative form
of delayed observability. When a weak perturbation couples an
otherwise invariant operator sector to its complement, the
resulting error in the selected expectation value depends on
which transitions between sectors can contribute to the readout.

Let $A(\varepsilon)=A_0+\varepsilon B$ describe a weakly perturbed
adjoint generator, where $A_0$ and $B$ are real skew-symmetric
matrices. Let $P$ be an orthogonal projector onto a sector
invariant under $A_0$, containing both the observable and the
input-state readout,
\begin{equation}
[A_0,P]=0,
\quad
Pw=w,
\quad
Ps=s.
\label{eq:weaksymmetryassumptions}
\end{equation}
We approximate the full dynamics by including the within-sector part of the perturbation,
\begin{equation}
A_P(\varepsilon)
:=
P(A_0+\varepsilon B)P.
\label{eq:projectedperturbedgenerator}
\end{equation}

\begin{proposition}[Absence of first-order readout error]
\label{prop:weaksymmetry}
Under \cref{eq:weaksymmetryassumptions}, the exact mean value
$f_\varepsilon(t)
=w^\top e^{tA(\varepsilon)}s$
and its projected approximation
$\widetilde f_\varepsilon(t)
=w^\top e^{tA_P(\varepsilon)}s$
satisfy
\begin{equation}
f_\varepsilon(t)-\widetilde f_\varepsilon(t)
=
O(\varepsilon^2)
\qquad
(\varepsilon\to0),
\label{eq:weaksymmetryerror}
\end{equation}
uniformly on every fixed compact time interval.
\end{proposition}

The result follows because first-order contributions involve only
the within-sector perturbation $PBP$, which is included in the
projected generator. Any omitted contribution must leave
this sector and return before reaching the readout, requiring
at least two perturbative couplings. In particular, if $PBP=0$,
the perturbation has no first-order effect on the mean value even
when it generates first-order components outside this sector.

For permutation symmetry, the invariant component of a
weak perturbation is included through $PBP$, while a purely
symmetry-breaking component has no coupling within the
permutation-invariant sector and is
therefore invisible at first order
\cite{Fulton.1991-RepresentationTheoryFirst}.
For translation
symmetry, the analogous separation is between zero- and nonzero-momentum
components. Contributions from the latter
can reach a translation-invariant readout only when their combined
momenta sum to zero modulo $n$.

\section{Numerical demonstrations}
\label{sec:Numerics}

We examine three aspects of approximate Heisenberg propagation.
First, we assess how accurately recorded defects can be used to
estimate the signed error in a selected expectation value.
Second, we compare approximate propagation in symmetry-adapted
coordinates with propagation in individual operator strings.
Finally, we test the first-order readout protection established
in \cref{prop:weaksymmetry}.

All readout errors are evaluated against an untruncated reference
for the corresponding target evolution. We summarize error magnitudes and fitted exponents by medians, binary comparisons by fractions, and sampling uncertainty by $95\%$ bootstrap confidence intervals.
Additional numerical demonstrations and implementation details
are provided in Appendix~\ref{app:ExtendedNumerics}.

\subsection{Signed transfer from recorded defects}
\label{subsec:NumericsDefectTransfer}

\begin{figure*}[htbp]
\centering
\includegraphics[width=0.9\textwidth]
{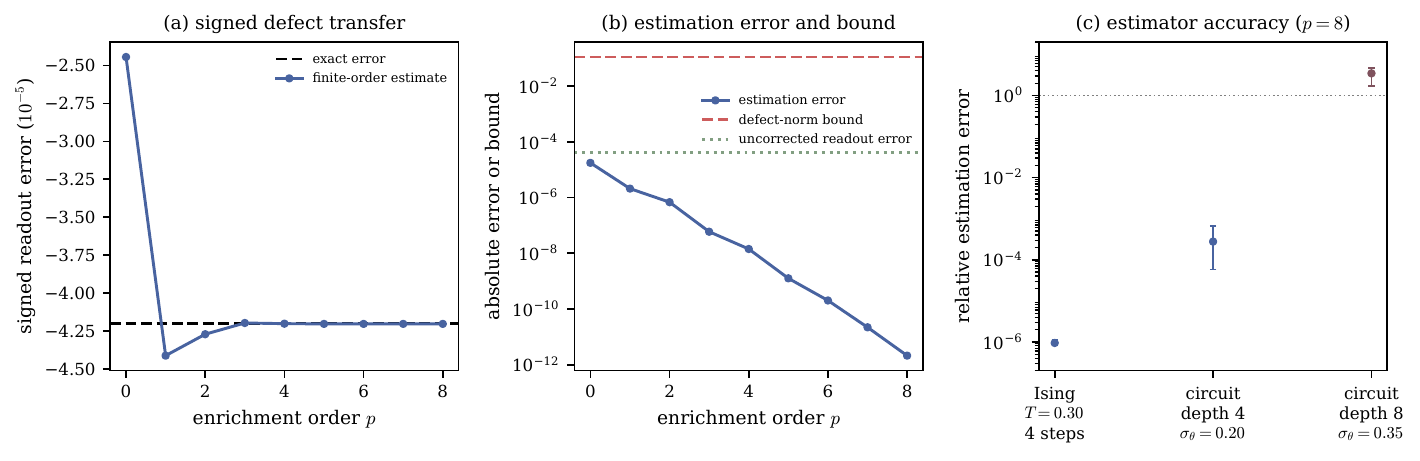}
\caption{
Signed readout-error estimates from discarded Pauli components.
(a) Exact signed error and finite-order transfer estimates
for $X/Y$-grade truncation of a transverse-field Ising circuit.
Every recorded defect has minimum $X/Y$ grade two, so its
transfer moments vanish below order $\Delta=2$.
The enrichment order $p$ counts additional transfer orders
beyond $\Delta$.
(b) Absolute estimation discrepancy compared with the
uncorrected readout error and the conservative defect-norm
bound for the same calculation.
(c) Relative estimation error $R^{(8)}$ from \cref{eq:numRelativeTransferError} for three ensembles
of $50$ six-qubit instances each, using an $X/Y$-grade
cutoff of one. The dotted line marks $R^{(8)}=1$. Points show medians and error bars denote
$95\%$ bootstrap confidence intervals from $5000$ resamples.
The mixed-field Ising ensemble uses $T=0.30$ and four
product-formula steps. The random-circuit ensembles have
depths four and eight, with single-qubit rotation-angle
standard deviations $\sigma_\theta=0.20$ and $0.35$,
respectively. Two-qubit rotation angles have standard
deviation $0.7\sigma_\theta$.
}
\label{fig:SignedDefectTransfer}
\end{figure*}

We first examine whether the discarded components recorded
during truncated Pauli propagation can be used to estimate the
resulting readout error. Following the layered defect formulation
of \cref{prop:layereddefecttransfer}, we record the Pauli terms
removed at each truncation step and evaluate their signed
contributions after propagation through the remaining circuit.

We consider the periodic transverse-field Ising Hamiltonian
\begin{equation}
H_\mathrm{TFI}
=
\sum_{j=1}^{n} Z_jZ_{j+1}
+
\sum_{j=1}^{n} X_j,
\qquad
Z_{n+1}\equiv Z_1,
\label{eq:numTFIM}
\end{equation}
with input state and observable
\begin{equation}
\rho_{\mathrm{in}}
=
\lvert0\rangle\!\langle0\rvert^{\otimes n},
\qquad
O=\frac{1}{n}\sum_{j=1}^{n}Z_j.
\label{eq:numInputObservable}
\end{equation}
For $n=6$ and total evolution time $T=0.2$, we use a first-order
product-formula circuit with four time steps of duration
$T/4$. 
After each elementary Pauli-rotation conjugation, we discard
strings whose $X/Y$ grade exceeds one, using \cref{eq:xygrade}.
The reference expectation value is computed by propagating
the same circuit without truncation. Thus, the measured
difference contains only the error introduced by discarding
Pauli components, with no contribution from the
product-formula approximation.

Every recorded defect has minimum grade two.
Since the adjoint action of each circuit generator lowers
the $X/Y$ grade by at most one,
\cref{thm:delayedobservability} implies that its transfer
moments vanish at orders zero and one.
The first allowed transfer order is therefore $\Delta=2$.

Figure~\ref{fig:SignedDefectTransfer}(a) compares the exact
signed error, obtained by summing the transferred defects
according to \cref{prop:layereddefecttransfer}, with the
finite-order estimate of \cref{eq:layeredenrichedestimate}.
The exact signed error is approximately
$-4.20\times10^{-5}$, while the estimation discrepancy
at $p=8$ is approximately $2.1\times10^{-12}$.

Panel~(b) compares this discrepancy with the uncorrected
absolute readout error and a conservative bound obtained
by multiplying the sum of normalized Hilbert--Schmidt defect norms
by the dual norm of the input-state readout
(see Appendix~\ref{app:numdetails}).
The latter is approximately $1.1\times10^{-1}$,
compared with the actual readout error of
$4.20\times10^{-5}$.
The signed transfer estimate captures the propagation and
cancellation of the discarded contributions, providing
readout-specific information absent from the norm bound
\cite{Shao.2025-CharacterizingPauliPropagation}.

To examine the limitations of the finite-order estimate,
\cref{fig:SignedDefectTransfer}(c) compares three ensembles at $p=8$. The median relative estimation
errors are approximately $9.4\times10^{-7}$ for the
mixed-field Ising instances and $2.8\times10^{-4}$ for
the depth-four random circuits. For the depth-eight circuits
with larger rotation angles, the median error exceeds unity.
These results illustrate that the accuracy of a fixed-order
transfer estimate depends on the remaining dynamics.
Vanishing low-order transfer moments therefore do not guarantee an accurate estimate.

Finally note that computing the additional transfer moments becomes more
expensive as $p$ increases. Indeed, when estimator work is included,
correction outperforms the best tested uncorrected
alternative at comparable generated work in only about
$21\%$ of Ising cases and $15\%$ of circuit cases
(see Appendix~\ref{app:EstimatorScope}).
Thus, finite-order transfer estimation is primarily
useful for diagnosing readout-specific truncation errors
and can provide accurate corrections when the remaining
dynamics permits.

\subsection{Propagation in symmetry-adapted coordinates}
\label{subsec:NumericsStructured}

\begin{figure*}[htbp]
\centering
\includegraphics[width=\textwidth]
{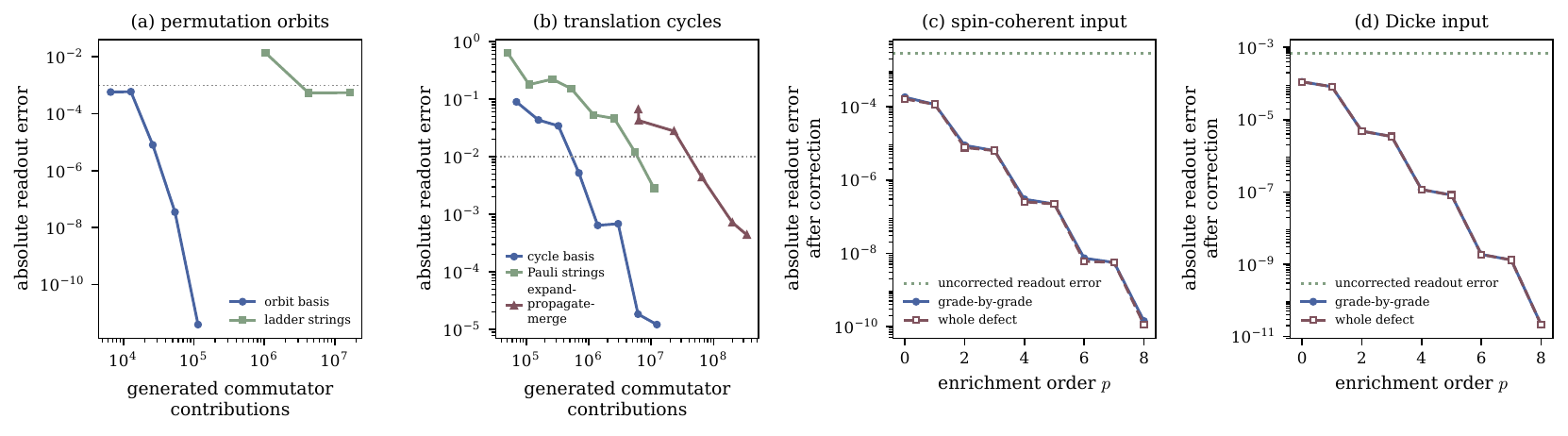}
\caption{
Approximate propagation and defect-transfer estimates in
symmetry-adapted operator coordinates.
(a) Direct permutation-orbit propagation and propagation
of individual ladder strings for the LMG model
\cref{eq:numLMG} at $n=12$ and $T=0.3$, using three first-order product-formula steps.
(b) Direct translation-cycle propagation, individual Pauli
strings, and merge-after symmetry compression for
\cref{eq:numTranslation} at $n=12$ and $T=1$, using eight first-order product-formula steps.
Both panels show absolute readout errors relative to the same untruncated circuit, plotted against generated commutator contributions. The dotted lines mark the comparison levels $10^{-3}$ in (a) and $10^{-2}$ in (b). Only completed calculations are shown.
(c) Absolute readout errors after correction using finite-order defect-transfer estimates for continuous projected evolution, with the spin-coherent input of \cref{eq:numCoherentState}. The grade-by-grade estimate treats components of each grade separately, while the whole-defect estimate uses a common expansion order.
(d) The same comparison for the Dicke input of \cref{eq:numDickeState} and observable of \cref{eq:numDickeObservable}. The two estimates coincide because the recorded defects have the same grade. In (c,d), dotted lines show the uncorrected readout errors.
}
\label{fig:StructuredCoordinates}
\end{figure*}

We now examine approximate propagation directly in
symmetry-adapted operator coordinates. Using the exact
permutation-orbit and translation-cycle representations of
Ref.~\cite{Barligea.2026-EnablingLieAlgebraicClassical},
we first compare the accuracy and computational work of
truncation in these representations with propagation of
individual operator strings. We then test the readout-aware
defect estimates in the symmetry-adapted representation.

\subsubsection{Permutation-orbit propagation}

For permutation-invariant dynamics, we represent observables
by sums of tensor-product operators related by permutations.
We use the ladder basis $\{I,Z,\sigma^+,\sigma^-\}$ introduced in \cref{subsec:StateAdaptedGradings}.
These ladder operators have definite weight under
commutation with $Z$,
$[Z,\sigma^\pm]=\pm2\sigma^\pm$.
This representation is convenient for the weight-based
readout gradings of \cref{subsec:StateAdaptedGradings}
and admits direct commutator rules in the orbit coordinates
\cite{Barligea.2026-EnablingLieAlgebraicClassical}.

For a product operator $Q$, let $\mathcal T_G(Q)$ denote
the set of its distinct images under a symmetry group $G$.
The corresponding orbit operator is
\begin{equation}
S^G_{[Q]}
:=
\sum_{Q'\in\mathcal T_G(Q)}Q',
\qquad
G\in\{S_n,C_n\},
\label{eq:orbitcyclesums}
\end{equation}
where $S_n$ acts by permutations and $C_n$ by cyclic
translations. In the permutation representation, each
orbit is specified by the multiplicities
$(n_Z,n_+,n_-)$, with $n_I=n-n_Z-n_+-n_-$.
The resulting invariant operator space contains
$\binom{n+3}{3}=\mathcal O(n^3)$ coordinates (cf.~Ref.~\cite{Albertini.2018-ControllabilitySymmetricSpin}).

We consider the Lipkin--Meshkov--Glick (LMG) model~\cite{Lipkin.1965-ValidityManybodyApproximation}
in its spin-$\tfrac12$ form with anisotropy $\gamma=0$~\cite{Dusuel.2004-FinitesizeScalingExponents},
\begin{equation}
H_{S_n}
=
\frac{1}{n}\sum_{i<j}X_iX_j
+
\frac{1}{2}\sum_i Z_i,
\label{eq:numLMG}
\end{equation}
with the input state and observable of
\cref{eq:numInputObservable}.

We compare direct orbit propagation with propagation
of the individual ladder strings represented by these
orbits. Both calculations use Top-$K$ truncation,
which selects at most $K$ operator components with the
largest coefficient magnitudes, accounting for their
basis norms, at each truncation step.
Varying $K$ controls the approximation and produces
the accuracy--work curves in
\cref{fig:StructuredCoordinates}(a).
The precise truncation procedure is described in
Appendix~\ref{app:StructuredNumericsMethods}.

For $n=12$ and $T=0.3$, we use a first-order
product-formula circuit with three time steps.
The reference is the same circuit propagated without
truncation. Computational work is measured by the
number of generated commutator contributions, counting
the target coordinates produced separately from each
source coordinate after combining contributions within
that source. This measure characterizes the propagation
work in the chosen representation and does not directly
determine execution time or memory usage.

The direct orbit calculation reaches a readout error
below $10^{-3}$ after approximately
$6.6\times10^3$ generated contributions.
The individual ladder-string calculation reaches
the same accuracy between approximately
$1.0\times10^6$ and $4.2\times10^6$ contributions.
At larger cutoffs, the orbit calculation agrees
with the untruncated reference to the numerical
accuracy shown.

\subsubsection{Translation-cycle propagation}

For translation-invariant dynamics, we use the
Pauli-cycle representation of
Ref.~\cite{Barligea.2026-EnablingLieAlgebraicClassical}.
We use averages of the distinct cyclic translates of each Pauli
string, obtained by dividing the orbit sums in
\cref{eq:orbitcyclesums} by their orbit sizes.
Their commutator action can be evaluated directly
without expanding the individual translates.
The number of independent cycles scales as
$\mathcal O(4^n/n)$, so the full invariant operator
space remains exponentially large.

We consider the periodic mixed-field Ising Hamiltonian
\begin{equation}
H_{C_n}
=
\sum_j Z_jZ_{j+1}
+
0.9\sum_j X_j
+
0.3\sum_j Z_j,
\label{eq:numTranslation}
\end{equation}
with the input state and observable of
\cref{eq:numInputObservable}.
For $n=12$ and $T=1$, we use eight first-order
product-formula steps. 

Figure~\ref{fig:StructuredCoordinates}(b) compares
three approaches: direct propagation and Top-$K$
truncation in cycle coordinates, propagation and
truncation of individual Pauli strings, and a
merge-after implementation that combines symmetry-related
contributions after string propagation, inspired by
\cite{Teng.2025-LeveragingSymmetryMerging} and specified in Appendix~\ref{app:StructuredMerging}.
All calculations are compared with the same
untruncated circuit.

Direct cycle propagation reaches a readout error
below $10^{-2}$ between approximately
$3.3\times10^5$ and $7.0\times10^5$
generated contributions. Individual Pauli-string propagation
reaches the same accuracy between approximately
$5.5\times10^6$ and $1.1\times10^7$ contributions.
The merge-after implementation additionally propagates
individual strings before combining symmetry-related
terms, increasing the work required in the tested regime.

Panels~(a,b) show that truncation performed directly
in symmetry-adapted coordinates can reach a given
readout accuracy with substantially fewer generated
commutator contributions. The comparisons also involve representation-dependent truncation schedules, specified in Appendix~\ref{app:StructuredMerging}.

\subsubsection{Defect transfer in structured coordinates}

We also examine whether defects generated during
approximate propagation in structured coordinates
admit the readout-aware estimates developed in
\cref{sec:Defects}. We consider continuous evolution
under \cref{eq:numLMG} at $n=6$ and $T=0.3$,
using two input states with different readout gradings.

Unlike the product-formula benchmarks of panels~(a,b),
these calculations use the continuously projected
dynamics of \cref{eq:projectedpropagation}.
The discarded components generate the time-dependent
boundary defect of the projected evolution, whose
contribution to the readout is given by
\cref{prop:continuousdefecttransfer}.
The corresponding transfer estimates are evaluated
using the continuous formulation in
Appendix~\ref{app:ReadoutEstimator}.

Figure~\ref{fig:StructuredCoordinates}(c) uses the
spin-coherent product state
\begin{equation}
\lvert\psi_{\mathrm{coh}}(\theta)\rangle
=
\left(
\cos\frac{\theta}{2}\,\lvert0\rangle
+
\sin\frac{\theta}{2}\,\lvert1\rangle
\right)^{\otimes n},
\quad
\theta=\frac{\pi}{3},
\label{eq:numCoherentState}
\end{equation}
with
$\rho_{\mathrm{in}}
=
\lvert\psi_{\mathrm{coh}}\rangle
\!\langle\psi_{\mathrm{coh}}\rvert$.
We use operator coordinates adapted to the rotated
Cartan algebra of this state, as described in
\cref{subsec:StateAdaptedGradings}, and restrict propagation to
components containing at most one ladder operator.
The measured observable is the mean magnetization
of \cref{eq:numInputObservable}.

The discarded components have different grades
and consequently different first allowed transfer
orders $\Delta$. We compare two prescriptions for
the finite-order transfer estimate.
The grade-by-grade estimate applies
\cref{eq:enrichedestimate} separately to each
homogeneous defect component, summing terms
through order $\Delta_g+p$ for grade $g$.
The whole-defect estimate uses the smallest
$\Delta$ among the discarded components and
includes all terms in the defect expansion through
that common order plus $p$.

Both estimates improve with $p$ in this example,
reducing the uncorrected readout error of approximately
$2.9\times10^{-3}$ to order $10^{-10}$ at $p=8$.
Their differences at intermediate orders reflect
the additional transfer orders included for
higher-grade components.

Figure~\ref{fig:StructuredCoordinates}(d) considers
the half-filled Dicke state
\begin{equation}
\lvert D_{n/2}^{(n)}\rangle
=
\binom{n}{n/2}^{-1/2}
\sum_{\substack{x\in\{0,1\}^n\\ |x|=n/2}}
\lvert x\rangle,
\label{eq:numDickeState}
\end{equation}
where $|x|$ denotes the Hamming weight of $x$ and
$\rho_{\mathrm{in}}
=
\lvert D_{n/2}^{(n)}\rangle
\!\langle D_{n/2}^{(n)}\rvert$.
The observable is
\begin{equation}
O_{\mathrm D}
=
\frac{\left(\sum_i Z_i\right)^2}{n^2}
-
\frac{I}{n}.
\label{eq:numDickeObservable}
\end{equation}

For this readout, we grade operators by their
absolute collective charge $|n_+-n_-|$ and restrict propagation to
components of grade at most two.
The recorded defects have a common grade, so
the grade-by-grade and whole-defect estimates coincide.
At $p=8$, the estimation discrepancy is approximately
$2.1\times10^{-11}$, compared with an uncorrected
readout error of approximately $7.0\times10^{-4}$.

These results demonstrate that signed defect-transfer
estimates can be evaluated directly in structured
operator coordinates, with the relevant grading
determined by the input-state readout.

\subsection{Protected readouts under weak symmetry breaking}
\label{subsec:NumericsWeakSymmetry}

\begin{figure*}[htbp]
\centering
\includegraphics[width=0.9\textwidth]
{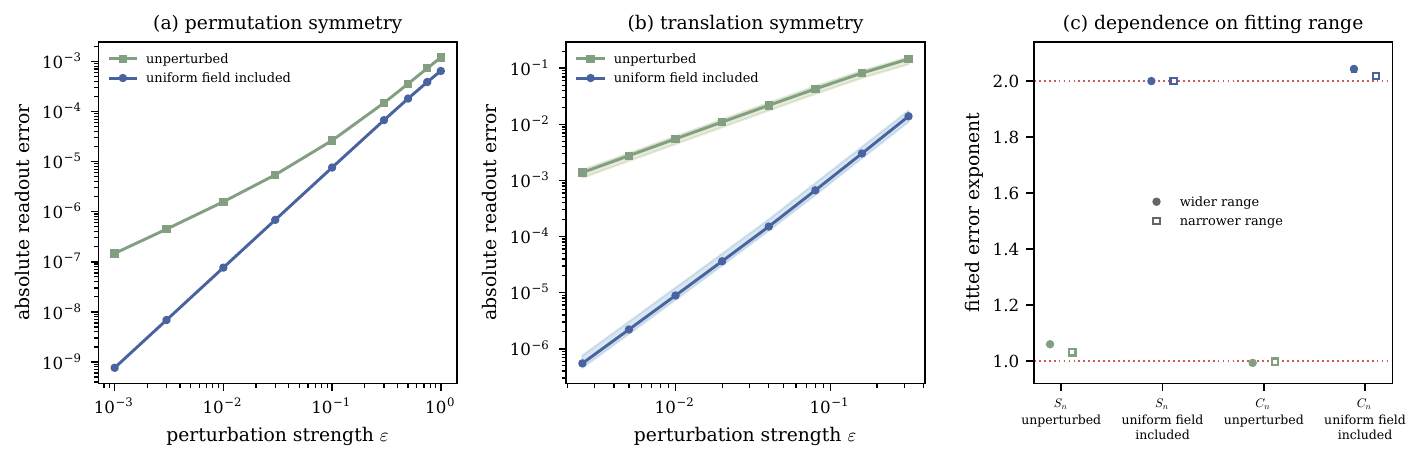}
\caption{
Readout errors under weak symmetry breaking.
(a) Permutation-invariant LMG dynamics and
(b) translation-invariant Ising dynamics with the
inhomogeneous field of \cref{eq:numWeakBreaking}.
The unperturbed approximation neglects the full perturbation,
while the approximation including the uniform field includes its
invariant component.
In (a,b), points show median absolute readout errors and shaded bands denote $95\%$ bootstrap confidence intervals over sampled inhomogeneous fields.
(c) Fitted error exponents for the two approximations
using the original and a shorter perturbative fitting
interval. The results approach linear scaling for the
unperturbed approximation and quadratic scaling for the
approximation including the uniform field, consistent with
\cref{prop:weaksymmetry}.
Error bars in (c) show $95\%$ bootstrap confidence intervals for the median fitted exponents.
}
\label{fig:ProtectedReadouts}
\end{figure*}

Finally, we test the first-order readout protection predicted
by \cref{prop:weaksymmetry}.
We perturb the permutation-invariant LMG Hamiltonian
\cref{eq:numLMG} and the translation-invariant Ising
Hamiltonian \cref{eq:numTranslation} with an inhomogeneous
longitudinal field,
\begin{equation}
H(\varepsilon)
=
H_0+\varepsilon\sum_j h_j Z_j,
\label{eq:numWeakBreaking}
\end{equation}
where $\varepsilon$ controls the perturbation strength.
The field can be decomposed into
\begin{equation}
\sum_j h_j Z_j
=
\overline h\sum_j Z_j
+
\sum_j(h_j-\overline h)Z_j,
\quad
\overline h=\frac{1}{n}\sum_j h_j.
\label{eq:numPerturbationSplit}
\end{equation}
The uniform component is invariant under both permutations
and cyclic translations. The remaining component has
vanishing symmetry average and does not act within the
corresponding invariant operator sector.

We compare unperturbed evolution under $H_0$ with evolution including the uniform field, generated by $H_0+\varepsilon\overline h\sum_j Z_j$. The figures label these approximations ``unperturbed'' and ``uniform field included'', respectively.
For the symmetric input state and observable of
\cref{eq:numInputObservable}, the latter satisfies the
conditions of \cref{prop:weaksymmetry}, which guarantees
a readout error of $O(\varepsilon^2)$ at fixed time.
The unperturbed approximation can have a first-order error.

\cref{fig:ProtectedReadouts}(a,b) show the absolute
readout errors for $n=12$ over independently sampled
inhomogeneous fields. Both symmetry settings exhibit
approximately linear scaling for the unperturbed approximation
and quadratic scaling for the
approximation including the uniform field at small $\varepsilon$.

\cref{fig:ProtectedReadouts}(c) the dependence
of the fitted power-law exponents on the perturbative fitting
interval.  For the narrower interval, the median exponents
are $1.03$ and $2.00$ for the permutation-invariant model,
and $1.00$ and $2.02$ for the translation-invariant model.
In each pair, the values correspond to the unperturbed
approximation and the approximation including the uniform
field, respectively. These results support the predicted
cancellation of the first-order readout error in both settings.
Further numerical details are given in
Appendix~\ref{app:WeakSymmetryNumerics}.

\section{Conclusion and outlook}
\label{sec:Conclusion}

We have developed a readout-aware theory of approximate Heisenberg
propagation based on weighted transfer from an observable source
to a state-defined readout on a commutator graph.
We call this basis-covariant formulation
\emph{generalized Pauli propagation}. It provides a common algebraic description of exact $\gsim$
\cite{Goh.2025-LiealgebraicClassicalSimulations},
Pauli propagation~\cite{Rudolph.2025-PauliPropagationComputational},
Majorana propagation~\cite{Miller.2025-SimulationFermionicCircuits},
symmetry-adapted representations
\cite{Barligea.2026-EnablingLieAlgebraicClassical},
and fixed-generator Krylov methods
\cite{Saad.1992-AnalysisKrylovSubspace},
while preserving their distinct computational structures. The framework separates the choice of operator representation
from the approximation applied to propagation and identifies
exact invariant-space reductions.

We derive exact identities expressing the mean-value error as
signed readout contributions from omitted source components and
defects generated during propagation
(see \cref{sec:Approximation,prop:continuousdefecttransfer,prop:layereddefecttransfer}).
Our central structural result, the delayed-observability theorem,
identifies conditions under which these contributions vanish
at low orders in propagation time
(see \cref{sec:Defects,thm:delayedobservability}).
If the readout vanishes on positive operator grades and one
commutator lowers the grade by at most
$\delta_{\mathrm{gr}}>0$, a defect of minimum grade $g$
has vanishing transfer moments below order
$\lceil g/\delta_{\mathrm{gr}}\rceil$.
We construct corresponding gradings for weight states and
generalized coherent states
(see \cref{subsec:StateAdaptedGradings}) and use them to
develop finite-order signed error estimates
(see \cref{subsec:ReadoutTransferEstimators}).
We further establish first-order readout protection under
weak symmetry breaking when the perturbation within the
invariant readout sector is included in the approximate generator
(see \cref{subsec:WeakSymmetry,prop:weaksymmetry}).

Our numerical demonstrations illustrate both the potential and
the practical limitations of this approach
(see \cref{sec:Numerics}).
Tracking the signed contributions of discarded Pauli components
can reveal readout errors that conservative norm bounds
substantially overestimate, although evaluating these contributions
can become computationally expensive
(see \cref{subsec:NumericsDefectTransfer}).
Working directly in symmetry-adapted coordinates can further
reduce propagation costs while allowing the same
readout-error analysis
(see \cref{subsec:NumericsStructured}).
Finally, the weak-symmetry experiments illustrate how the
structure of the input state and observable can suppress
approximation errors (see \cref{subsec:NumericsWeakSymmetry}).

Several extensions follow naturally. The same bilinear adjoint structure
extends to more general linear readouts, including real-time correlators and
parameter derivatives, for which related Pauli-propagation methods have
recently been developed
\cite{Kemper.2026-EfficientComputationRealtime,
Lin.2026-BackpropagatingPauliPropagation}.
The commutator-graph perspective may further help identify
reduced operator spaces adapted to both the dynamics and
the desired readout, with potential applications to
state-adapted projection methods
\cite{Perez.2026-StateAdaptedGeneralizedMeanField}
and approximate shadow Hamiltonian simulation
\cite{Somma.2025-ShadowHamiltonianSimulation,
Chakraborty.2026-EfficientAlgorithmApproximateShadow}.

Together, these results connect the algebraic structure and
computational representation of Heisenberg propagation to the
error in the expectation value being computed. They provide a common
readout-aware framework for assessing and refining approximate
simulation methods beyond exactly tractable regimes. In particular, this perspective can guide the development of
truncation strategies that prioritize computational effort
according to the contributions of discarded operator components
to the desired expectation value.

\section*{Code and Data availability}
The code and numerical data underlying the results presented in this
work, including the simulation parameters and configurations needed
to reproduce the numerical experiments, are available from the
corresponding author upon reasonable request.

\section*{Acknowledgments}
The authors thank Erik Schultheis for his careful review of the
manuscript and valuable suggestions that substantially improved its
quality. We also thank Matthew L.~Sims-Goh for helpful comments
on this work.
This research was carried out in part within the RISE Germany programme of the
German Academic Exchange Service (DAAD). Y.B.\ acknowledges support through
RISE Germany, which is funded by the German Federal Foreign Office.
J.S.K.\ and A.B.\ acknowledge funding from the German Federal Ministry of Research,
Technology and Space (BMFTR) under grant 13N17231 (HoliQC2), and support from
the Munich Quantum Valley, funded by the Bavarian State Government via the Hightech Agenda Bavaria. Numerical calculations were performed on
local computing resources and on the LiCCA high-performance computing cluster
of the University of Augsburg, co-funded by the Deutsche
Forschungsgemeinschaft (DFG, German Research Foundation), Project ID
499211671.

\bibliography{references}

\onecolumngrid
\appendix

\section{Operator coordinates and exact reductions}
\label{app:CoordinateExtensions}

\subsection{Nonorthogonal and complex operator coordinates}
\label{app:NonorthogonalCoordinates}

The adjoint representation in \cref{eq:adjointmatrix} assumes an
orthonormal Hermitian operator basis, so its matrix elements and
observable coefficients follow directly from Hilbert--Schmidt
inner products. For a linearly independent Hermitian basis
$\mathcal B=\{B_\alpha\}_{\alpha=1}^{D}$ that need not be
orthonormal, we account for the basis overlaps through the
Gram matrix
\begin{equation}
G_{\alpha\beta}
=
\langle B_\alpha,B_\beta\rangle_\kappa.
\end{equation}
Defining the adjoint overlaps by
${C_{\alpha\beta}
=\langle B_\beta,\operatorname{ad}_H(B_\alpha)\rangle_\kappa}$
and using the row-action convention
${\operatorname{ad}_H(B_\alpha)=\sum_\beta A_{\alpha\beta}B_\beta}$,
we obtain
\begin{equation}
C=AG,
\qquad
A=CG^{-1}.
\label{eq:nonorthogonaladjointmatrix}
\end{equation}
Skew-adjointness of $\operatorname{ad}_H$ implies
$C^\top=-C$, which gives
\begin{equation}
AG+GA^\top=0.
\label{eq:metricskewadjoint}
\end{equation}
The inverse Gram matrix also determines the observable expansion
coefficients. Specifically, the dual basis
\begin{equation}
\widetilde B^\alpha
=
\sum_\beta (G^{-1})_{\alpha\beta}B_\beta
\end{equation}
satisfies
$\langle\widetilde B^\alpha,B_\beta\rangle_\kappa
=\delta^\alpha_{\ \beta}$.
It therefore extracts the coefficients of an operator expansion
and the adjoint action according to
\begin{equation}
w_\alpha
=
\langle\widetilde B^\alpha,O\rangle_\kappa,
\qquad
A_{\alpha\beta}
=
\langle\widetilde B^\beta,\operatorname{ad}_H(B_\alpha)\rangle_\kappa.
\end{equation}
The input coordinates remain
$s_\alpha=\tr(B_\alpha\rho_{\mathrm{in}})$ as in
\cref{eq:observableandinputcoordinates}.
Consequently, the mean-value expressions of
\cref{eq:adjointmeanvalue,eq:layeredadjointmeanvalue}
remain valid without orthonormality.

The ladder operators $\sigma^\pm=(X\pm iY)/2$ used in
\cref{subsec:NumericsStructured} additionally require complex
operator coordinates. Since these operators are not Hermitian,
we use the sesquilinear Hilbert--Schmidt inner product
$\langle X,Y\rangle_\kappa=\kappa\tr(X^\dagger Y)$,
which is conjugate-linear in its first argument and defines
a positive Hilbert--Schmidt norm. For a linearly independent complex
basis $\mathcal B_{\mathbb C}=\{B_a\}_{a=1}^{D}$, the Gram matrix
becomes
\begin{equation}
G_{ab}
=
\kappa\tr(B_a^\dagger B_b).
\label{eq:complexgram}
\end{equation}
Writing $X=\sum_a x_aB_a$ and $s_a=\tr(B_a\rho)$,
the expectation value is still given by the bilinear contraction
$\tr(X\rho)=x^\top s$.
The coefficient dynamics obey $\dot x=A^\top x$, and
skew-adjointness now gives
\begin{equation}
(A^\top)^\dagger G+GA^\top=0.
\label{eq:complexmetricskewadjoint}
\end{equation}
Thus, $e^{tA^\top}$ preserves the norm induced by $G$,
while the exact readout is
$x^\top e^{tA}s$.

The metric also determines the norms entering the global
error bound of \cref{eq:globaldefectbound}. In complex
coordinates, the Hilbert--Schmidt norm and its dual readout norm are
\begin{equation}
\lVert X\rVert_\kappa^2
=
x^\dagger Gx,
\qquad
\lVert s\rVert_{*,G}^2
=
s^\top G^{-1}\overline{s}.
\label{eq:complexreadoutnorm}
\end{equation}
Indeed, writing
$x^\top s=(G^{-1}\overline{s})^\dagger Gx$,
the Cauchy--Schwarz inequality gives
$|x^\top s|\leq\lVert X\rVert_\kappa\lVert s\rVert_{*,G}$.
Applying this inequality to the exact defect-transfer identity
\cref{eq:meanvaluedefecttransfer} yields
\begin{equation}
|\epsilon_{\mathrm{mv}}(t)|
\leq
\lVert s\rVert_{*,G}
\left(
\lVert q_0\rVert_G
+
\int_0^t\lVert d(\tau)\rVert_G\,\mathrm d\tau
\right),
\end{equation}
where $\lVert x\rVert_G^2=x^\dagger Gx$.
For an orthonormal Hermitian basis, $G=I$, recovering
\cref{eq:globaldefectbound}.

Finally, the metric affects coordinate truncation.
An idempotent coordinate projector $P$ is orthogonal with
respect to $G$ precisely when $P^\dagger G=GP$.
Otherwise, coordinate deletion generally defines an
oblique projection. The source and boundary defects of
\cref{eq:projecteddefect} and their exact transfer remain
valid, since these identities require only $P^2=P$.
For an orthogonal but unnormalized basis, individual
coefficient norms are $|x_a|\sqrt{G_{aa}}$.
For a nonorthogonal basis, cross terms in $x^\dagger Gx$
prevent an independent coordinatewise norm decomposition.
Hence, changes of operator coordinates preserve exact
propagation but can alter graph sparsity, coefficient-based
truncation, and computational cost.

\subsection{Proof of the reachable-and-observable reduction}
\label{prf:reachableobservableproof}

\begin{proof}[Proof of \cref{prop:reachableobservable}]
For any $v\in\mathcal N_{\mathcal G}(s)$, generator $A_j$,
and generator word $(j_1,\ldots,j_m)$,
\begin{equation}
    (A_j^\top v)^\top A_{j_1}\cdots A_{j_m}s
    =
    v^\top A_jA_{j_1}\cdots A_{j_m}s
    =0.
\end{equation}
Thus $\mathcal N_{\mathcal G}(s)$ is invariant under every
$A_j^\top$. The empty-word condition further gives
$\mathcal N_{\mathcal G}(s)\subseteq\ker(s^\top)$.

To establish maximality, let
$\mathcal W\subseteq\ker(s^\top)$ be any subspace invariant
under every $A_j^\top$. For $v\in\mathcal W$ and any
generator word, invariance implies
$A_{j_m}^\top\cdots A_{j_1}^\top v\in\mathcal W$.
Consequently,
\begin{equation}
    v^\top A_{j_1}\cdots A_{j_m}s
    =
    (A_{j_m}^\top\cdots A_{j_1}^\top v)^\top s
    =0,
\end{equation}
which proves $\mathcal W\subseteq\mathcal N_{\mathcal G}(s)$.

Finally, expanding the layer exponentials into their
convergent power series shows that
\begin{equation}
    v^\top e^{\theta_LA_L}\cdots e^{\theta_1A_1}s=0
\end{equation}
for every $v\in\mathcal N_{\mathcal G}(s)$ and every
circuit generated by $\mathcal G$.
Hence, replacing $w$ by $w+v$ leaves all corresponding
mean values unchanged. Restricting this equivalence
to the observable-reachable module yields the quotient
in \cref{eq:reachableobservablequotient}.
\end{proof}

\section{Defect-transfer identities and approximation mechanisms}
\label{app:DefectTransfer}

\subsection{Proofs of the continuous and discrete defect-transfer identities}
\label{app:defecttransferproof}

\begin{proof}[Proof of \cref{prop:continuousdefecttransfer}]
The exact coefficient trajectory satisfies
$\dot w(t)=A^\top w(t)$. By \cref{eq:defectdefinition},
the coefficient error $\epsilon_w(t)=w(t)-\widetilde w(t)$
therefore obeys
\begin{equation}
\dot{\epsilon}_w(t)
=
A^\top\epsilon_w(t)+d(t),
\qquad
\epsilon_w(0)=q_0.
\end{equation}
Applying the variation-of-constants
formula~\cite{Hochbruck.2010-ExponentialIntegrators} gives
\begin{equation}
\epsilon_w(t)
=
e^{tA^\top}q_0
+
\int_0^t e^{(t-\tau)A^\top}d(\tau)\,\mathrm d\tau,
\end{equation}
which establishes \cref{eq:defectduhamel}.
Contracting with the input expectation vector $s$ yields
\cref{eq:meanvaluedefecttransfer}, since
$\epsilon_{\mathrm{mv}}(t)=\epsilon_w(t)^\top s$.
\end{proof}

\begin{proof}[Proof of \cref{prop:layereddefecttransfer}]
By \cref{eq:layeroperatorerror,eq:generallayerdefect},
the operator error satisfies
\begin{equation}
\epsilon_{O,\ell-1}
=
\mathcal E_\ell(\epsilon_{O,\ell})+D_{\ell-1}.
\label{eq:layererrorrecursion}
\end{equation}
Starting from $\epsilon_{O,L}=Q_L$ and iterating this
recursion for $\ell=L,\ldots,1$ gives
\begin{equation}
\epsilon_{O,0}
=
\mathcal E_1\cdots\mathcal E_L(Q_L)
+
\sum_{c=0}^{L-1}
\mathcal E_1\cdots\mathcal E_c(D_c),
\end{equation}
where the product for $c=0$ is the identity.
This establishes \cref{eq:layertelescoping}.
Taking the expectation value with respect to
$\rho_{\mathrm{in}}$ and using linearity of the trace
then yields \cref{eq:layerdefecttransfer}.
\end{proof}

\subsection{Proof of the source and boundary defect formulas}
\label{app:sourceboundarydefectsproof}

\begin{proof}[Proof of \cref{cor:projecteddefect}]
For the projected evolution of \cref{eq:projectedpropagation},
the initial condition satisfies $\widetilde w(0)=Pw$.
Since $P^2=P$ and the generator $PA^\top P$ preserves
$\operatorname{ran}P$, we have
$P\widetilde w(t)=\widetilde w(t)$ for all $t$.
Using \cref{eq:defectdefinition}, the dynamic defect becomes
\begin{equation}
d(t)
=
A^\top\widetilde w(t)-PA^\top P\widetilde w(t)
=
(I-P)A^\top\widetilde w(t).
\end{equation}
The initial condition also gives the source error
\begin{equation}
q_0=w-\widetilde w(0)=(I-P)w,
\end{equation}
which establishes \cref{cor:projecteddefect}.
\end{proof}

\subsection{Projected propagation and invariant-sector cuts}
\label{app:sharedirreps}

When the projected subspace is invariant under the dynamics,
projected propagation generates no boundary defect.
In particular, if the projector $P$ commutes with the
coefficient generator $A^\top$, then
\begin{equation}
d(t)=(I-P)A^\top\widetilde w(t)=0.
\end{equation}
The exact error identity of \cref{eq:invarianttail}
therefore reduces to the propagated source tail
\begin{equation}
\epsilon_{\mathrm{mv}}(t)
=
w^\top(I-P)^\top e^{tA}s.
\end{equation}
Since $P^\top$ commutes with $A$, this contribution
vanishes whenever $(I-P)^\top s=0$.

Invariant-sector cuts also arise from decomposing the
operator space into irreducible representations of the
compact group generated by the dynamics, as considered
in Ref.~\cite{Monbroussou.2026-ClassicalSimulationModel}.
To relate this construction to the reachable-and-observable
reduction of \cref{prop:reachableobservable}, consider a
multiplicity-free decomposition of the complexified operator
space,
\begin{equation}
\mathcal V_{\mathbb C}
=
\bigoplus_{k\in\mathcal I}\mathcal V_k,
\end{equation}
where $\mathcal I$ indexes the pairwise inequivalent
irreducible invariant subspaces $\mathcal V_k$.
Let $P_k$ denote the orthogonal projector onto
$\mathcal V_k$, and define $w_k=P_kw$ and $s_k=P_ks$.
A nonzero observable component generates its entire
irreducible sector under the group action, while a sector
with $s_k=0$ is unobservable to the readout.
The reachable-and-observable quotient is therefore
isomorphic to the shared-support space
\begin{equation}
 \mathcal V_{\mathrm{shared}}
:=
\bigoplus_{\substack{k\in\mathcal I\\w_k\neq0,\ s_k\neq0}}
\mathcal V_k.
\label{eq:sharedirrepsupport}
\end{equation}
Thus, only sectors shared by the observable source and
the state readout contribute to the exact mean value.
For representations with repeated irreducible components,
the quotient of \cref{prop:reachableobservable} remains
the applicable reduction, since irrep labels alone do
not resolve the observable and unobservable directions
within multiplicity spaces.

\subsection{Summary of approximation mechanisms and their defects}
\label{app:TruncationAtlas}

Approximate operator-propagation methods differ in how they restrict
the represented operator space, discard generated components, or
modify the dynamics. The defect-transfer identities of
\cref{prop:continuousdefecttransfer,prop:layereddefecttransfer}
apply to these mechanisms once their discrepancies are recorded
relative to the exact propagation.
\Cref{tab:truncationatlas} summarizes representative methods and
their error channels. We distinguish initial source omissions
($\mathrm S$), dynamically discarded components ($\mathrm D$),
modified generators ($\mathrm M$), and finite-Krylov-depth errors
($\mathrm K$). Exact reductions introduce no error ($0$).
A source error occurs only when the approximation modifies the
initial observable.

\begin{table}[t]
\centering
\caption{
Representative operator-propagation mechanisms, their error
channels, and method-specific qualifications.
The channel labels identify the origin of an approximation
relative to the exact target dynamics.
}
\label{tab:truncationatlas}
\scriptsize
\setlength{\tabcolsep}{3pt}
\renewcommand{\arraystretch}{1.15}
\begin{tabular}{
@{}
p{0.15\linewidth}|
p{0.24\linewidth}
p{0.24\linewidth}
p{0.31\linewidth}
@{}
}
\toprule
\textbf{Mechanism}
& \textbf{Representative methods}
& \textbf{Defect source}
& \textbf{Control or qualification}
\\
\midrule

\raggedcell{Exact invariant reductions}
&
\raggedcell{Structured-basis propagation
\cite{Barligea.2026-EnablingLieAlgebraicClassical}
and symmetry merging
\cite{Teng.2025-LeveragingSymmetryMerging}}
&
\raggedcell{$0$ for exact reductions without pruning.}
&
Requires the relevant invariance and representation conditions.
Representative merging need not preserve arbitrary operators.
\\ \hline

\raggedcell{Initial invariant-sector cuts}
&
Irreducible-sector truncation
\cite{Monbroussou.2026-ClassicalSimulationModel}
&
$\mathrm S$ from omitted invariant sectors.
No dynamic defect.
&
Exact dynamics within each selected sector.
Available error guarantees depend on the representation
and averaging assumptions.
\\ \hline

\raggedcell{Dynamic coordinate cuts}
&
\raggedcell{Pauli-weight truncation and related approximations
\cite{Rudolph.2025-PauliPropagationComputational,
Schuster.2024-PolynomialtimeClassicalAlgorithm,
Angrisani.2025-ClassicallyEstimatingObservables,
Angrisani.2026-SimulatingQuantumCircuits,
Aharonov.2023-PolynomialTimeClassicalAlgorithm}.
Coefficient thresholds and Top-$K$
\cite{Begusic.2025-RealTimeOperatorEvolution,
Shao.2025-CharacterizingPauliPropagation}.
Spatial diameter
\cite{HoldenDye.2026-DiameterTruncatedOperator}
and Majorana degree
\cite{Miller.2025-SimulationFermionicCircuits}.}
&
$\mathrm D$ from discarded components,
with additional $\mathrm S$ if the initial observable is truncated.
&
Noise-, locality-, complexity-, or model-dependent guarantees.
Adaptive cuts require the actual discarded contribution
at each step.
\\ \hline

\raggedcell{Metric-dependent projections}
&
\raggedcell{State-adapted covariance projection
\cite{Perez.2026-StateAdaptedGeneralizedMeanField}}
&
$\mathrm S$ and $\mathrm D$ when initial or propagated
components are omitted.
&
Projection and norm estimates depend on the operator metric.
State-dependent approximations require their actual residual.
\\ \hline

\raggedcell{Generator modifications}
&
\raggedcell{Dissipation-assisted operator evolution (DAOE)
\cite{Rakovszky.2020-DissipationassistedOperatorEvolution}}
&
$\mathrm M$ from the difference between the exact
and modified generators.
&
Guarantees for the modified dynamics do not directly
bound their discrepancy from the original dynamics.
\\ \hline

\raggedcell{Finite-dimensional exponential-action approximations}
&
Krylov methods
\cite{Saad.1992-AnalysisKrylovSubspace,
Hochbruck.1997-KrylovSubspaceApproximations,
Jawecki.2020-ComputableUpperError,
Matsuura.2026-KrylovBreakTimes}
&
$\mathrm K$ from finite Krylov depth.
&
Residual estimates for exact matrix-vector products.
Inexact products require additional control.
\\

\bottomrule
\end{tabular}
\end{table}

Generator modifications contribute to the full-space dynamic defect
even when no operator coordinates are discarded. If the approximate
trajectory satisfies
$\dot{\widetilde w}(t)=\widetilde A^\top\widetilde w(t)$,
then \cref{eq:defectdefinition} gives
$d(t)
=
(A^\top-\widetilde A^\top)\widetilde w(t)$.
For coordinate truncations, the defect instead records components
removed during propagation. This also applies to nonlinear truncation
rules such as adaptive Top-$K$, where the layer discrepancy must
be evaluated from the computed approximate trajectory.
For nonorthogonal coordinates, the corresponding projection and
norm conventions follow \cref{app:NonorthogonalCoordinates}.

Finite Krylov depth introduces a distinct approximation to the
exponential action within a fixed operator space. Standard Krylov
residual estimates assume exact matrix-vector products.
Approximating these products introduces additional errors that
require separate treatment
\cite{Dinh.2017-AnalysisInexactKrylov}.
In particular, a Krylov estimate for the projected generator does
not control operator components discarded before or during
the propagation.

The defect-transfer identities describe errors in the propagated
operator. Approximating their subsequent transfer introduces a
separate estimator-order error without modifying the operator
trajectory. Stochastic propagation and measurement introduce
sampling uncertainty
\cite{Rall.2019-SimulationQubitQuantum,
Dougherty.2026-UnifiedFrameworkOperator}.
Product-formula error arises when a layered circuit approximates
a target continuous evolution, while floating-point and
numerical-reference errors affect the computed results.
These contributions must be distinguished when assessing the
total error, without counting a discrepancy already included
in the full-space defect a second time.

\section{Readout gradings and delayed observability}
\label{app:DelayedObservability}

\subsection{Delayed observability and generator words}
\label{app:delayedobsproof}

\begin{proof}[Proof of \cref{thm:delayedobservability}]
By \cref{eq:transfermoments}, the transfer moments expand into
weighted commutator-graph walks,
\begin{equation}
r^\top A^m s
=
\sum_{\alpha_0,\ldots,\alpha_m}
r_{\alpha_0}
A_{\alpha_0\alpha_1}\cdots
A_{\alpha_{m-1}\alpha_m}s_{\alpha_m},
\end{equation}
where the product of matrix elements is unity for $m=0$.
For each nonzero contribution, $r_{\alpha_0}\neq0$ and
$A_{\alpha_{j-1}\alpha_j}\neq0$ for $j=1,\ldots,m$.
Repeated application of \cref{eq:gradeloweringcondition} gives
\begin{equation}
\operatorname{gr}(B_{\alpha_m})
\geq
\operatorname{gr}_{\min}(r)-m\delta_{\mathrm{gr}}.
\end{equation}
For $\delta_{\mathrm{gr}}>0$ and $m<\Delta(r)$, the
right-hand side is strictly positive. The readout condition
\cref{eq:gradereadoutcondition} therefore implies
$s_{\alpha_m}=0$. Every contribution vanishes, proving
\cref{eq:forbiddenmoments}. Substituting these vanishing moments
into \cref{eq:transfermoments} yields
\cref{eq:delayedtransferexpansion}.

For a sequence of generators $A_{j_1},\ldots,A_{j_m}$,
the corresponding transfer moment is
\begin{equation}
r^\top A_{j_1}\cdots A_{j_m}s
=
\sum_{\alpha_0,\ldots,\alpha_m}
r_{\alpha_0}
\left(
\prod_{k=1}^m (A_{j_k})_{\alpha_{k-1}\alpha_k}
\right)s_{\alpha_m}.
\end{equation}
If every generator satisfies \cref{eq:gradeloweringcondition}
with the same $\delta_{\mathrm{gr}}$, each nonzero product
obeys the same grade bound. Hence all scalar transfer moments associated with generator words of
length $m<\Delta(r)$ vanish, establishing
\cref{eq:forbiddengeneratorwords}.

If $\delta_{\mathrm{gr}}=0$, the grade cannot decrease
under any generator. For
$\operatorname{gr}_{\min}(r)>0$, every reachable component
therefore has positive grade, so all transfer moments
vanish and $\Phi_r(t)=0$ for every $t$.
If $\operatorname{gr}_{\min}(r)=0$, the grading imposes
no vanishing condition on the transfer moments.
\end{proof}

\subsubsection{Finite-word unobservable spaces}
\label{app:FiniteWordUnobservability}

These vanishing transfer moments define a finite-order analogue
of the exact unobservable space in \cref{prop:reachableobservable}. For $M\geq1$, let
\begin{equation}
\mathcal N_{\mathcal G}^{(<M)}(s)
:=
\left\{
v:\,
v^\top A_{j_1}\cdots A_{j_m}s=0,\;
0\leq m<M
\right\},
\label{eq:finitewordunobservable}
\end{equation}
where the condition holds for every generator word,
including the empty word. These spaces satisfy
$\mathcal N_{\mathcal G}^{(<M+1)}(s)
\subseteq\mathcal N_{\mathcal G}^{(<M)}(s)$, and
\begin{equation}
\mathcal N_{\mathcal G}(s)
=
\bigcap_{M\geq1}
\mathcal N_{\mathcal G}^{(<M)}(s).
\label{eq:finitewordintersection}
\end{equation}
For $\delta_{\mathrm{gr}}>0$ and $\Delta(r)\geq1$,
\cref{eq:forbiddengeneratorwords} implies
$r\in\mathcal N_{\mathcal G}^{(<\Delta(r))}(s)$.
For $\delta_{\mathrm{gr}}=0$ and
$\operatorname{gr}_{\min}(r)>0$, the defect belongs
to the exact unobservable space $\mathcal N_{\mathcal G}(s)$.

Unlike $\mathcal N_{\mathcal G}(s)$, finite-word spaces
need not be invariant under the dual generator actions
$A_j^\top$. They therefore certify vanishing transfer
only up to the specified word length and cannot generally
be removed through an exact quotient reduction.

\subsection{Proofs of weight selection and bounded grade jumps}
\label{app:WeightSelectionProofs}

\begin{proof}[Proof of \cref{prop:cartanreadout}]
For $X_\nu\in\mathcal V_\nu$, cyclicity of the trace and
\cref{eq:cartandiagonalstate,eq:operatorweightdecomposition} give
\begin{equation}
\nu_k\tr(\rho_{\mathrm{in}}X_\nu)
=
\tr([\rho_{\mathrm{in}},h_k]X_\nu)
=
0.
\end{equation}
For every nonzero weight $\nu$, at least one component
$\nu_k$ is nonzero. Hence
$\tr(\rho_{\mathrm{in}}X_\nu)=0$, proving
\cref{eq:weightselectionrule}.
\end{proof}

\begin{proof}[Proof of \cref{lem:weightshift}]
For $X_\nu\in\mathcal V_\nu$ and a Hamiltonian component
$H_\mu$ of definite weight, the Jacobi identity and
\cref{eq:operatorweightdecomposition,eq:hamiltonianweights}
give
\begin{equation}
[h_k,[H_\mu,X_\nu]]
=
(\mu_k+\nu_k)[H_\mu,X_\nu].
\end{equation}
Since the commutator action of $H_\mu$ preserves
$\mathcal V_{\mathbb C}$, every nonzero
$[H_\mu,X_\nu]$ belongs to $\mathcal V_{\mu+\nu}$.
By \cref{eq:weightlengthcondition,eq:weightgrade},
its grade satisfies
\begin{equation}
\ell(\nu+\mu)
\geq
\ell(\nu)-\ell(\mu)
\geq
\ell(\nu)-\delta_{\mathrm{gr}}.
\end{equation}
Summing over the Hamiltonian components in
\cref{eq:hamiltonianweights} establishes the same
grade-lowering bound for $\operatorname{ad}_H$
in a basis respecting the weight grading.
\end{proof}

\subsection{Pauli, coherent-state, and Dicke readouts}
\label{app:CartanPrimer}

The Cartan generators $h_k$ of \cref{subsec:StateAdaptedGradings}
admit simultaneous eigenstates
$h_k\lvert\lambda\rangle=\lambda_k\lvert\lambda\rangle$,
called weight states~\cite{Hall.2015-LieGroupsLie}.
Their projectors commute with every $h_k$ and therefore
satisfy the readout selection rule of \cref{prop:cartanreadout}.
The corresponding operator-weight decomposition
\cref{eq:operatorweightdecomposition} requires the commutator
actions of the Cartan generators to preserve
$\mathcal V_{\mathbb C}$.
Since $X_\nu^\dagger\in\mathcal V_{-\nu}$, the condition
$\ell(-\nu)=\ell(\nu)$ in \cref{eq:weightlengthcondition}
allows opposite-weight components to form Hermitian operators
within the same grade. A coordinate transformation that mixes
different grades need not preserve the grading conditions of
\cref{thm:delayedobservability}.

\paragraph{All-zero-state readout.}
\label{app:PauliWeightSpecialization}

For the input state $\lvert0\rangle^{\otimes n}$, consider the
local Cartan algebra
$\mathfrak h=\operatorname{span}\{Z_1,\ldots,Z_n\}$.
The weight-adapted single-qubit basis
$\{I,Z,\sigma^+,\sigma^-\}$ satisfies
\begin{equation}
\sigma^\pm=\frac{X\pm iY}{2},
\qquad
[Z,\sigma^\pm]=\pm2\sigma^\pm.
\end{equation}
Each tensor-product operator has a weight vector
$\nu\in\{0,\pm2\}^n$. Choosing
$
\ell(\nu)=\frac{1}{2}\|\nu\|_1
$
assigns a grade equal to its number of ladder operators.
The Hermitian Pauli operators $X$ and $Y$ are combinations
of opposite-weight ladder operators with equal grade.
Their tensor products therefore recover the $X/Y$ grading
of \cref{eq:xygrade}. Positive-grade components have
vanishing expectation in $\lvert0\rangle^{\otimes n}$.

\paragraph{Generalized coherent-state readout.}

A generalized coherent state (GCS) is obtained by acting
with the associated Lie group on an extremal weight
state~\cite{Perelomov.1972-CoherentStatesArbitrary,
Gilmore.1972-GeometrySymmetrizedStates,
Perelomov.1986-GeneralizedCoherentStates,
Zhang.1990-CoherentStatesTheory}.
For $\lvert\psi\rangle=U\lvert\lambda\rangle$,
the density operator
$\rho=U\lvert\lambda\rangle\!\langle\lambda\rvert U^\dagger$
commutes with the rotated Cartan algebra
$\mathfrak h_U=U\mathfrak h U^\dagger$.
Consequently, \cref{prop:cartanreadout} applies to operator
weights defined relative to $\mathfrak h_U$, provided
the corresponding Cartan commutator actions preserve the
chosen operator space. Equivalently, conjugating the
Hamiltonian and observable by $U^\dagger$ restores the
original weight decomposition
\cite{Somma.2005-QuantumComputationComplexity,
Somma.2006-EfficientSolvabilityHamiltonians}.
Neither construction alone guarantees efficient access
to the resulting operator coordinates or state overlaps.

The spin-coherent input of \cref{eq:numCoherentState}
is obtained from $\lvert0\rangle^{\otimes n}$ by the
collective local rotation
\begin{equation}
U_\theta
=
\bigotimes_{j=1}^n e^{-i\theta Y_j/2},
\qquad
\theta=\frac{\pi}{3}.
\end{equation}
We use the rotated Cartan generators
$U_\theta Z_jU_\theta^\dagger$ and the corresponding
ladder operators to define the readout grading.
Equivalently, we propagate the transformed Hamiltonian
$U_\theta^\dagger H U_\theta$ and observable
$U_\theta^\dagger O U_\theta$ in the unrotated basis.
We establish the weight grading in the full tensor-product
operator space, where the local Cartan actions preserve
the representation. Permutation symmetry then allows
us to combine operators of equal grade into homogeneous
orbit sums. The resulting grading applies directly to
the structured propagation of
\cref{fig:StructuredCoordinates}(c).

\paragraph{Dicke-state readout.}

The half-filled Dicke state of \cref{eq:numDickeState}
is an eigenstate of the collective Cartan generator
$\sum_j Z_j$. It therefore satisfies the selection rule
for collective operator weights, although its projector
does not generally commute with the individual $Z_j$.
For a ladder orbit labeled by $(n_Z,n_+,n_-)$, the
collective weight is $2(n_+-n_-)$.
We consequently use the grade
\begin{equation}
\operatorname{gr}_{\mathrm D}(n_Z,n_+,n_-)
=
|n_+-n_-|.
\end{equation}
Unlike the ladder degree $n_++n_-$, this grade vanishes
for operators containing equal numbers of raising and
lowering factors. Such operators can contribute directly
to the Dicke-state readout even at nonzero ladder degree.
A non-extremal Dicke state is a collective-spin weight
state but is not generally a spin-coherent GCS
\cite{Perelomov.1986-GeneralizedCoherentStates}.
The readouts in \cref{fig:StructuredCoordinates}(c,d)
therefore require different gradings.

Finally, a readout grading does not by itself control
the number of operator coordinates used in propagation.
Let $\mathcal V^X_{\leq W}$ denote the span of Pauli
strings with $X/Y$ grade at most $W$. Its dimension is
\begin{equation}
\dim\mathcal V^X_{\leq W}
=
2^n\sum_{w=0}^{W}\binom{n}{w}.
\end{equation}
Even the grade-zero sector contains $2^n$ operators.
For example, $Z^{\otimes n}$ has $X/Y$ grade zero
but Pauli support weight $n$.
The grading therefore constrains the first permitted
readout-transfer order without guaranteeing efficient
propagation or justifying low-support truncation.

\section{Finite-order readout-transfer estimates}
\label{app:ReadoutEstimator}

\subsection{Continuous defect convolution}
\label{app:ContinuousDefectEstimator}

The dynamic contribution to the mean-value error in
\cref{eq:meanvaluedefecttransfer} depends on the defect
at each propagation time. Suppose it admits a decomposition
\begin{equation}
d(t)=\sum_j y_j(t)f_j,
\label{eq:defectdecomposition}
\end{equation}
where the defect directions $f_j$ are time-independent
and the amplitudes $y_j(t)$ carry the time dependence.
Linearity of the transfer function~\eqref{eq:defecttransferfunction} gives
\begin{equation}
\epsilon_{\mathrm{mv,dyn}}(t)
=
\sum_j\int_0^t
y_j(\tau)\Phi_{f_j}(t-\tau)\,\mathrm d\tau.
\label{eq:dynamictransferfunctions}
\end{equation}
Each defect is therefore weighted by its amplitude at
the time of generation and propagated over the remaining
interval $t-\tau$.

For defect directions with finite certified delay
$\Delta(f_j)$, substituting the enriched expansion~\eqref{eq:enrichedestimate} yields
\begin{equation}
\begin{aligned}
\widehat\epsilon_{\mathrm{mv,dyn}}^{(p)}(t)
=
\sum_j\sum_{m=\Delta(f_j)}^{\Delta(f_j)+p}
\mu_m(f_j)
\ \times
\int_0^t y_j(\tau)
\frac{(t-\tau)^m}{m!}\,\mathrm d\tau.
\end{aligned}
\label{eq:continuousenrichedestimate}
\end{equation}
Including the initial source error gives
\begin{equation}
\widehat\epsilon_{\mathrm{mv}}^{(p)}(t)
=
\widehat\Phi_{q_0}^{(p)}(t)
+
\widehat\epsilon_{\mathrm{mv,dyn}}^{(p)}(t).
\label{eq:continuousdefectestimate}
\end{equation}
Exactly unobservable defect directions contribute zero
and may be omitted from these expansions.

Decomposing a defect into components of a single grade is optional. In the
whole-defect prescription, each recorded defect direction
$f_j$ receives a single delay $\Delta(f_j)$ determined
by its minimum supported grade. Alternatively, writing
$f_j=\sum_g f_{j,g}$ with homogeneous components
$f_{j,g}$ permits separate cutoffs
$\Delta(f_{j,g})+p$. Both prescriptions estimate the
transfer of the same approximate trajectory, but can
include different total orders and require different
computational work. Neither has a general accuracy
advantage at fixed $p$.

For exponential amplitudes, the convolution can be
evaluated analytically through the functions
\begin{equation}
\varphi_k(z)
:=
\int_0^1
e^{(1-\theta)z}
\frac{\theta^{k-1}}{(k-1)!}\,\mathrm d\theta.
\label{eq:phifunction}
\end{equation}
In particular,
\begin{equation}
\int_0^t
e^{\lambda\tau}
\frac{(t-\tau)^m}{m!}\,\mathrm d\tau
=
t^{m+1}\varphi_{m+1}(t\lambda).
\label{eq:phiconvolution}
\end{equation}
These functions are standard in exponential integrators
\cite{Hochbruck.2010-ExponentialIntegrators}.

\subsection{Ordered layered estimates}
\label{app:LayeredDefectEstimator}

For layered propagation, the transfer of a recorded defect
$d_c$ follows the ordered product
\cref{eq:ordereddefecttransfer}. Expanding each exponential
and including total word lengths from $\Delta(d_c)$
through $\Delta(d_c)+p$ gives
\begin{equation}
\begin{aligned}
\widehat\Phi_c^{(p)}(d_c)
=
\sum_{\substack{
m_1,\ldots,m_c\geq0\\
\Delta(d_c)\leq|\boldsymbol m|\\
|\boldsymbol m|\leq\Delta(d_c)+p
}}
\left(
\prod_{\ell=1}^{c}
\frac{\theta_\ell^{m_\ell}}{m_\ell!}
\right)
\ \times\
d_c^\top A_c^{m_c}\cdots A_1^{m_1}s,
\end{aligned}
\label{eq:layeredenrichedestimate}
\end{equation}
where $|\boldsymbol m|=\sum_{\ell=1}^{c}m_\ell$.
The generator order is inherited from the remaining
circuit and requires no commutativity assumption.

If every generator satisfies the common grade-lowering
condition, \cref{eq:forbiddengeneratorwords} excludes
all terms with $|\boldsymbol m|<\Delta(d_c)$.
Higher-order terms may also vanish through additional
selection rules or cancellations. The cutoff therefore
specifies the next $p$ Taylor orders beyond the first allowed transfer order, without requiring their coefficients to be nonzero.

Let $q_L$ denote the coefficient vector of the initial
source error $Q_L$ in \cref{eq:layertelescoping}. The
complete signed estimate is
\begin{equation}
\widehat\epsilon_{\mathrm{mv}}^{(p)}
=
\widehat\Phi_L^{(p)}(q_L)
+
\sum_{c=0}^{L-1}
\widehat\Phi_c^{(p)}(d_c).
\end{equation}
The source term propagates through all $L$ layers.
For $c=0$, the remaining product is the identity,
so the exact local contribution is
$\Phi_0(d_0)=d_0^\top s$.

\subsection{Remainders and numerical residuals}
\label{app:EstimatorRemainder}

For a defect $r$ with finite $\Delta(r)$, the remainder
of \cref{eq:enrichedestimate} contains no terms below order
$\Delta(r)+p+1$. In orthonormal operator coordinates,
submultiplicativity and the Cauchy--Schwarz inequality give
\begin{equation}
\begin{aligned}
\bigl|\Phi_r(t)-\widehat\Phi_r^{(p)}(t)\bigr|
\leq{}
\|r\|_2\|s\|_2
\ \times
\sum_{m=\Delta(r)+p+1}^{\infty}
\frac{(|t|\|A\|_2)^m}{m!}.
\end{aligned}
\label{eq:estimatorremainderbound}
\end{equation}
The corresponding metric-dependent formulation follows
from the Hilbert--Schmidt and dual readout norms of
\cref{app:NonorthogonalCoordinates}.
For continuous defects, this bound applies inside
\cref{eq:dynamictransferfunctions}, weighted by
$|y_j(\tau)|$ and evaluated over the remaining time
$t-\tau$. The source and dynamic remainder bounds
can then be combined by the triangle inequality.

An analogous bound applies to the ordered layered
estimator without assuming that the generators commute.
Define
\begin{equation}
a_c
:=
\sum_{\ell=1}^{c}
|\theta_\ell|\,\|A_\ell\|_2.
\end{equation}
Bounding each generator word and summing over all
multi-indices of fixed total degree gives
\begin{equation}
\begin{aligned}
\bigl|\Phi_c(d_c)-\widehat\Phi_c^{(p)}(d_c)\bigr|
\leq{}
\|d_c\|_2\|s\|_2
\ \times
\sum_{m=\Delta(d_c)+p+1}^{\infty}
\frac{a_c^m}{m!}.
\end{aligned}
\end{equation}
Here the multinomial identity converts the sum of
absolute contributions at total degree $m$ into
$a_c^m/m!$. The same bound applies to the initial
source error with $c=L$.

These estimates control the discrepancy between a
defect's exact transfer and its finite-order estimate.
The exponential series
converges for every finite propagation time, although
the remainder bound can be conservative at low order.

A separate residual arises when propagation within a
selected operator space is evaluated approximately.
Let $V$ embed the coordinates of that space into the full
space, let $P$ be the corresponding projector with
$PV=V$, and let $y_K(t)$ denote a numerical
exponential-action approximation. Its full-space
dynamic defect separates as
\begin{equation}
\begin{aligned}
A^\top Vy_K-V\dot y_K
={}
(I-P)A^\top Vy_K
+
\bigl(PA^\top Vy_K-V\dot y_K\bigr).
\end{aligned}
\end{equation}
The first term records boundary leakage from
that operator space. The second is the numerical
propagation residual within that space. Standard Krylov
residual estimates control the latter for exact
matrix-vector products
\cite{Saad.1992-AnalysisKrylovSubspace,
Jawecki.2020-ComputableUpperError}.
When these products are themselves approximated,
additional inexact-Krylov errors must be considered
\cite{Dinh.2017-AnalysisInexactKrylov}.

The total readout error consequently depends on both
the recorded propagation defects and the accuracy of
their subsequent transfer estimates.

\section{Numerical methods and estimator limitations}
\label{app:ExtendedNumerics}

We provide implementation details and additional tests for the
numerical demonstrations of \cref{sec:Numerics}.
For product-formula experiments, truncation errors are
measured against the untruncated evolution of the same
ordered circuit. Continuous-time experiments use the
corresponding Hamiltonian evolution as their reference.
The following subsections specify the truncation procedures,
numerical approximations, and statistical methods relevant
to each experiment.

\subsection{Pauli defect-transfer experiments}
\label{app:numdetails}

\paragraph{Product-formula benchmark.}

The deterministic calculation in
\cref{fig:SignedDefectTransfer}(a,b) uses the periodic
transverse-field Ising Hamiltonian of \cref{eq:numTFIM}
with $n=6$, the all-zero input state, and the intensive
magnetization observable of \cref{eq:numInputObservable}.
We split $H_{\mathrm{TFI}}=H_{ZZ}+H_X$, where
$H_{ZZ}=\sum_j Z_jZ_{j+1}$ and $H_X=\sum_j X_j$.
The circuit consists of four first-order steps with
$\Delta t=0.05$. In the Schrödinger convention,
\begin{equation}
U_{\mathrm{PF}}=(e^{-i\Delta t H_{ZZ}}e^{-i\Delta t H_X})^4.
\label{eq:numTFIProductFormula}
\end{equation}
In Heisenberg application order, each step conjugates
the observable by the $ZZ$ rotations followed by the
$X$ rotations. The $ZZ$ bonds are applied in the order
$(1,2)$, $(2,3)$, $(3,4)$, $(4,5)$, $(6,1)$, $(5,6)$,
followed by $X_1,\ldots,X_6$.
The reference uses the same ordered rotations without
operator truncation.

After each elementary conjugation, we discard Pauli
strings with $X/Y$ grade greater than one. This gives
48 truncation opportunities, of which 18 produce
nonzero recorded defects. Each defect record collects
all components discarded at that truncation step.
The implementation additionally removes coefficients
below $10^{-15}$ as numerical cleanup.
Consequently, the recorded-defect identities are
verified up to floating-point accuracy.

Writing $D_c=\sum_P d_{c,P}P$ for the defect recorded
at step $c$, its exact contribution is propagated
through the remaining circuit according to
\cref{prop:layereddefecttransfer}.
The initial observable is not truncated, so no
source error is introduced. Every nonzero recorded
defect has minimum $X/Y$ grade two, giving
$\Delta=2$ under the applicable generator-word bound.
The exact signed error is approximately
$-4.20\times10^{-5}$, while the discrepancy of the
$p=8$ transfer estimate is approximately
$2.1\times10^{-12}$.

\paragraph{Conservative defect-norm bound.}

The bound plotted in \cref{fig:SignedDefectTransfer}(b)
uses normalized Hilbert--Schmidt norms.
For unnormalized Pauli strings,
$\|D_c\|_\kappa^2=\sum_P|d_{c,P}|^2$ with
$\kappa=2^{-n}$.
The all-zero-state readout vector has Euclidean norm
$\|s\|_2=2^{n/2}$. Unitary propagation preserves the
Hilbert--Schmidt norm, so the Cauchy--Schwarz inequality
and \cref{eq:layerdefecttransfer} give
\begin{equation}
|\epsilon_{\mathrm{mv}}|\leq B_{\mathrm{HS}}:=2^{n/2}\sum_c\left(\sum_P|d_{c,P}|^2\right)^{1/2}.
\label{eq:numDefectNormBound}
\end{equation}
For this calculation,
$B_{\mathrm{HS}}\approx1.11\times10^{-1}$.
The bound does not account for cancellations between
transferred defects or their alignment with the readout.

The truncated propagation generates 679 units of work,
while the $p=8$ estimator requires 251399 additional
units, giving 252078 in total. Here computational work counts generated anticommuting
Pauli-rotation branches, excluding the
untruncated reference calculation.

\paragraph{Ensemble transfer-estimation tests.}

For \cref{fig:SignedDefectTransfer}(c), we evaluate
the relative discrepancy of the finite-order estimate,
\begin{equation}
R^{(p)}=\frac{|\widehat\epsilon_{\mathrm{mv}}^{(p)}-\epsilon_{\mathrm{mv}}|}{|\epsilon_{\mathrm{mv}}|}.
\label{eq:numRelativeTransferError}
\end{equation}
All three ensembles contain 50 independently generated
six-qubit instances and use the all-zero input state,
intensive $Z$ magnetization, an $X/Y$-grade cutoff
of one, and enrichment order $p=8$.
We exclude instances with
$|\epsilon_{\mathrm{mv}}|\leq10^{-12}$
from relative-error and sign statistics.
None of the three displayed ensembles contains such
an exclusion.

The mixed-field Ising ensemble uses
$H=\sum_j Z_jZ_{j+1}+0.9\sum_j X_j+\sum_j h_jZ_j$,
with independent $h_j\sim\mathcal U[-0.6,0.6]$.
We use $T=0.30$ and four product-formula steps.
Each Pauli string is encoded by two integers $(x,z)$ whose bits indicate
the sites carrying $X/Y$ and $Z/Y$ factors, respectively, with site one
occupying the least significant bit. We sort terms lexicographically
by $(x,z)$ and assign each term to the first existing group with whose
members it commutes, creating a new group if necessary.
The elementary rotations are applied in group-creation order and
in sorted term order within each group.

The random-circuit ensembles contain four or eight
blocks with single-qubit rotation-angle standard
deviations $\sigma_\theta=0.20$ and $0.35$,
respectively. Each block applies ascending
single-site Pauli rotations followed by alternating
nearest-neighbor bonds on an open chain.
Pauli letters are sampled uniformly from
$\{X,Y,Z\}$, and bond angles have standard
deviation $0.7\sigma_\theta$.
The two regimes reuse the same realization seeds
but differ in depth and angle scale.

The median relative discrepancies are approximately
$9.4\times10^{-7}$ for mixed-field Ising,
$2.8\times10^{-4}$ for the depth-four circuits,
and $3.47$ for the depth-eight circuits.
The final ensemble demonstrates that vanishing
low-order transfer moments does not guarantee
an accurate finite-order estimate at larger
rotation angles.

\subsection{Structured-coordinate propagation and estimation}
\label{app:StructuredNumericsMethods}

\paragraph{Coordinate normalization and Top-$K$ truncation.}

The comparisons in \cref{fig:StructuredCoordinates}(a,b)
use Top-$K$ truncation in different operator representations.
Since a structured coordinate represents an entire symmetry
orbit, its coefficient magnitude alone does not measure
the norm of the corresponding operator contribution.
For permutation-adapted ladder coordinates, we use
unnormalized orbit sums $B_a=\sum_{Q\in\mathcal O_a}Q$,
labeled by $a=(n_Z,n_+,n_-)$, with metric
\begin{equation}
G_{aa}=|\mathcal O_a|\,2^{-(n_++n_-)}.
\label{eq:ladderorbitmetric}
\end{equation}
For translation symmetry, we instead use cycle averages
$B_{[Q]}=m_{[Q]}^{-1}\sum_{Q'\in[Q]}Q'$,
where $m_{[Q]}$ is the number of distinct translates.
These have $G_{[Q][Q]}=1/m_{[Q]}$ and span the same
space as the cycle sums of \cref{eq:orbitcyclesums},
with $x_{\mathrm{avg}}=m_{[Q]}x_{\mathrm{sum}}$.

To account for these different normalizations,
Top-$K$ selects the coordinates with the largest
Hilbert--Schmidt contributions,
\begin{equation}
|x_a|\sqrt{G_{aa}},\qquad G_{aa}=2^{-n}\tr(B_a^\dagger B_a).
\label{eq:numTopKScore}
\end{equation}
Ties are resolved by coordinate-key order.
For complex ladder coordinates, Top-$K$ truncation
does not necessarily select both members of a
Hermitian-conjugate orbit pair. We therefore report
the real part of the resulting expectation value.

We compare the resulting approximation errors against
the work required to generate commutator contributions.
This counts distinct target coordinates produced by each
source coordinate, after combining its contributions
to the same target. The measure captures the work of
expanding operators in each representation.

\paragraph{Product-formula benchmarks.}

The experiments in \cref{fig:StructuredCoordinates}(a,b)
compare approximate propagation in symmetry-adapted
coordinates with propagation of individual operator
strings. Both use the same symmetry-invariant dynamics,
so the untruncated representations describe the same
operator evolution. Truncation changes which components
are propagated and how much work is required to propagate
them. We therefore compare readout accuracy against
generated commutator work, using the untruncated
evolution of the same product-formula circuit as the
reference. This excludes product-formula error from
the reported discrepancies.

For the permutation-invariant LMG model
\cref{eq:numLMG}, each step applies the collective
$XX$ interaction followed by the longitudinal field
in Heisenberg order. For the translation-invariant
Ising model \cref{eq:numTranslation}, the $ZZ$ and
longitudinal-field terms precede the transverse field.
The corresponding group exponentials are evaluated
using adaptively subdivided Taylor expansions.

The truncation schedule also affects the comparison.
All methods truncate individual Taylor contributions,
but the permutation-orbit implementation additionally
truncates the accumulated operator after each Hamiltonian
group, whereas the ladder-string baseline applies its
accumulated-operator cutoff only after complete steps.
In the translation comparison, direct cycle and
Pauli-string propagation use the same group- and
step-level truncation points.

\paragraph{Expand--propagate--merge baseline.}
\label{app:StructuredMerging}

To assess the benefit of propagating directly in symmetry-adapted
coordinates, \cref{fig:StructuredCoordinates}(b) also includes an
expand--propagate--merge baseline. This approach expands the selected
cycle operators into individual Pauli strings, propagates them through
a complete product-formula step, and recombines symmetry-related
contributions before applying Top-$K$ truncation. Unlike direct
structured propagation
\cite{Barligea.2026-EnablingLieAlgebraicClassical},
it therefore performs the propagation in the larger
Pauli-string representation.

Expansion and recombination introduce no approximation by
themselves. Let $V$ embed structured coefficient vectors into individual-string
coordinates, and let $V^\sharp$ be a left inverse satisfying $V^\sharp V=I$.
If an exact layer update $\mathsf F_\ell$ preserves
the invariant operator space, its structured representation
satisfies
$
V^\sharp\mathsf F_\ell V=\overline{\mathsf F}_\ell$.
Consequently, both approaches produce identical results when
their exact updates and subsequent truncations are matched.
In the experiment, however, merge-after truncates only after
complete product-formula steps, whereas direct cycle propagation
also truncates intermediate Taylor contributions and Hamiltonian
groups. The comparison therefore reflects both the coordinate
representation and the truncation schedule.

Our expand--propagate--merge baseline is inspired by the
symmetry-merging approach of
Ref.~\cite{Teng.2025-LeveragingSymmetryMerging}.
While that method propagates canonical Pauli representatives
and combines symmetry-related descendants, we explicitly
expand cycle operators into individual Pauli strings and
merge them after each complete product-formula step.
This provides a complementary comparison between direct
propagation in structured coordinates and propagation
with subsequent symmetry reduction.
The merge-after calculations at $K=1600$ and $K=3200$
exceeded the limit of $5\times10^5$ expanded Pauli
coordinates and are therefore omitted from \cref{fig:StructuredCoordinates}.

\paragraph{Continuous projected propagation.}

The experiments in \cref{fig:StructuredCoordinates}(c,d)
instead examine whether recorded defects can accurately
predict the readout error of a continuously projected
evolution. We use the LMG model at $n=6$ and $T=0.3$,
with a fixed grade projector applied during Lanczos
propagation. The discarded commutator contributions
determine the time-dependent boundary defect of
\cref{eq:projecteddefect}, whose effect on the readout
is estimated through the convolution in
\cref{eq:continuousenrichedestimate}.
Here, no product-formula approximation or Top-$K$ truncation
is involved. Lanczos propagation used a maximum dimension of 120
and a breakdown tolerance of $10^{-10}$.
The projected calculations terminated at dimensions
13 and 15 for the coherent and Dicke inputs, respectively.
As a separate reference check, independent dense calculations
agreed with the untruncated structured expectation values
to within $7\times10^{-16}$.

The grading is chosen according to the input-state
readout. For the coherent state of
\cref{eq:numCoherentState}, we restrict propagation to components with ladder degree
at most one in rotated coordinates. The discarded
components have degrees two and three, corresponding
to first allowed transfer orders of one and two,
respectively. For the Dicke state of
\cref{eq:numDickeState}, we instead restrict propagation to components with absolute
collective charge at most two. The nonzero discarded
components have charge four and a common delay
$\Delta=2$.

These different defect structures explain the
comparison between whole-defect and grade-by-grade
transfer estimation. In the coherent-state experiment,
separate grade-dependent cutoffs include additional
transfer orders for the higher-grade components.
At $p=8$, this increases estimator work from
97594 to 194610 generated contributions. For the Dicke
readout, all recorded defects have the same grade,
so both prescriptions coincide and require 24130
generated contributions.

\subsection{Weak-symmetry ensembles and fitting procedure}
\label{app:WeakSymmetryNumerics}

The experiments in \cref{fig:ProtectedReadouts} test whether
including the symmetry-invariant part of a weak perturbation
removes the first-order readout error predicted by
\cref{prop:weaksymmetry}. We compare continuous state-vector
evolution under the full Hamiltonian with the unperturbed
approximation and approximations including the uniform field of
\cref{subsec:NumericsWeakSymmetry}. For each symmetry
setting, we use $n=12$ and sample 24 independent
inhomogeneous longitudinal fields. We use $T=0.3$ for the permutation-invariant LMG model
and $T=1$ for the translation-invariant Ising model. The corresponding
absolute errors are
\begin{equation}
E_{\mathrm{unpert}}(\varepsilon)=|f_\varepsilon-f_0|,\qquad E_{\mathrm{uniform}}(\varepsilon)=|f_\varepsilon-f_{\mathrm{uniform},\varepsilon}|.
\label{eq:numWeakSymmetryErrors}
\end{equation}

The sampled fields contain a nonzero uniform component,
so neglecting the entire perturbation generally produces
a first-order error. For permutation symmetry, we generate
independent $z_j\sim\mathcal N(0,1)$ and set
\begin{equation}
h_j=\frac12+\frac{z_j-\overline z}{\sqrt{n^{-1}\sum_k(z_k-\overline z)^2}}.
\end{equation}
This fixes the sample mean at $1/2$ and the mean-square
inhomogeneity at one. For translation symmetry, we use
$h_j=1+z_j$, with independent standard Gaussian $z_j$.
In both cases, the approximation including the uniform field selects
the actual uniform component $\overline h\sum_j Z_j$.
Each field realization is reused across perturbation
strengths and both approximations.

To assess the perturbative scaling, we fit
$
E(\varepsilon)\propto\varepsilon^{\alpha_{\mathrm{fit}}}
$
separately for each realization by linear regression
in logarithmic coordinates, using positive errors above
$10^{-14}$. We compare the original and shorter fitting
windows, with upper bounds $0.03$ and $0.01$ for
permutation symmetry and $0.08$ and $0.02$ for
translation symmetry. The reported exponents are
medians of the individual fitted slopes, with $95\%$
confidence intervals obtained from $5000$ bootstrap
resamples of the field realizations.

\subsection{Reliability and total work of defect correction}
\label{app:EstimatorScope}

\begin{figure}[t]
    \centering
    \includegraphics[width=\linewidth]{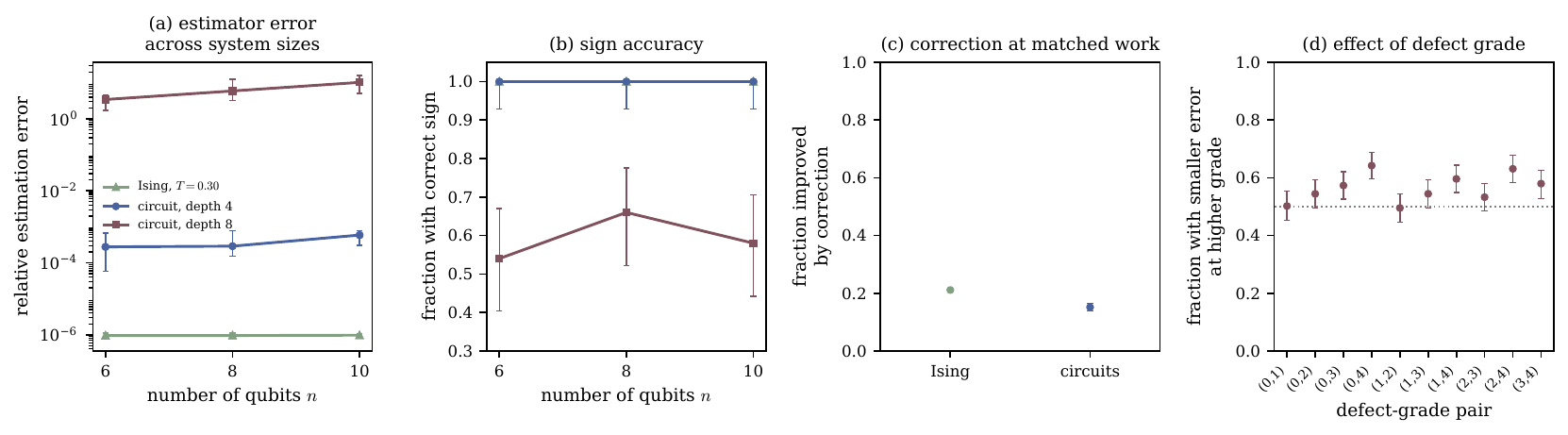}
    \caption{
    Reliability and computational cost of defect correction.
    (a) Relative estimation error and
    (b) fraction of correctly predicted error signs for the
    grade-truncated Ising and random-circuit ensembles at
    $n=6,8,10$. Short-time Ising instances are excluded
    because their errors fall below the numerical threshold.
    The ensembles use the parameters specified in \cref{fig:SignedDefectTransfer}(c), with enrichment order $p=8$.
    (c) Fraction of corrected trajectories that improve upon
    the best tested uncorrected trajectory at no greater total
    generated work, including estimator cost.
    (d) Fraction of matched defect pairs for which the
    higher-grade defect has a smaller estimation error,
    at fixed total expansion order eight.
    Error bars show pointwise $95\%$ Wilson binomial confidence
intervals in (b) and $95\%$ bootstrap confidence intervals
in (a,c,d).
    }
    \label{fig:EstimatorScope}
\end{figure}

The examples of \cref{subsec:NumericsDefectTransfer} show that
recorded defects can provide accurate estimates of signed readout
errors. We now examine how reliably this accuracy persists across
different propagation regimes and whether the improvement justifies
the additional computational work.

Figures~\ref{fig:EstimatorScope}(a,b) extend the ensemble tests
of \cref{app:numdetails} to $n=6,8,10$, using the
$X/Y$-grade cutoff $W=1$ and enrichment order $p=8$.
Each ensemble contains 50 independent realizations per system size.
The mixed-field Ising and depth-four circuit ensembles exhibit small relative estimation errors and accurately predicted signs.
In contrast, the depth-eight circuits exhibit relative errors exceeding
unity and substantially less reliable sign predictions.
Short-time Ising instances are excluded because their exact
readout errors fall below the numerical resolution threshold
$10^{-12}$. These differences illustrate that the vanishing
moments established by \cref{thm:delayedobservability}
do not determine the magnitude of the subsequent transfer terms.

Accuracy alone, however, does not establish a computational
advantage. To assess the cost of defect correction,
\cref{fig:EstimatorScope}(c) considers a broader set of
21 truncation policies, including $X/Y$-grade cutoffs $W=0,\ldots,3$,
support and diameter cutoffs $1,\ldots,4$, Top-$K$
with $K\in\{n,2n,4n\}$, combinations of grade-two
truncation and Top-$K$, and DAOE thresholds
$\ell_*=1,2,3$ with damping strength one and
amplitude cutoff $10^{-4}$
(cf.~\cref{app:TruncationAtlas}). The comparison covers both
propagation regimes and three observables: the intensive magnetization of
\cref{eq:numInputObservable}, the local operator $Z_1$,
and the nearest-neighbor average
$(n-1)^{-1}\sum_{j=1}^{n-1}Z_jZ_{j+1}$. The Ising comparisons include the short-time setting
$T=0.12$ with two product-formula steps and the
intermediate-time setting $T=0.30$ with four steps.
For each instance and observable, we compare the corrected
readout with the most accurate tested uncorrected trajectory
whose propagation work does not exceed the corrected
trajectory's total propagation and estimator work.
The correction improves upon this reference in approximately
$21\%$ of the resolved Ising comparisons and $15\%$ of the
circuit comparisons. These fractions pool 15757 resolved
Ising comparisons and 17633 circuit comparisons, drawn
from 150 independent realization blocks per model family.
Confidence intervals in \cref{fig:EstimatorScope}(c,d)
use 5000 bootstrap resamples of realization blocks,
stratified by system size. All comparisons sharing a
realization are resampled together. Thus, although defect correction can
substantially improve a given approximate trajectory,
it does not generally outperform alternative truncation
choices at comparable generated work.

Finally, \cref{fig:EstimatorScope}(d) tests whether higher-grade
defects are intrinsically easier to estimate.
We compare synthetic defects of equal norm propagated through
the same remaining circuit, fixing the total Taylor expansion
order at eight. Higher-grade defects yield smaller estimation
errors in approximately $56\%$ of the comparisons.
The grade therefore determines which low-order transfer
moments vanish, but does not generally order the accuracy
of finite-order estimates.

\end{document}